\documentclass[11pt,a4paper]{article}

\newcommand{\citemsugraandcmssm}{\cite{Chamseddine:1982jx,Barbieri:1982eh,Ibanez:1982ee,Hall:1983iz,Ohta:1982wn,Kane:1993td}}
\newcommand{\citemttwo}{\cite{Lester:1999tx,Barr:2003rg,Cheng:2008hk}}

\newcommand{\meff}{\ensuremath{m_{\mathrm{eff}}}} 

\def\ptmiss{\ensuremath{\vec{p}_\mathrm{T}^{\mathrm{miss}}}}

\usepackage{jheppub}
\usepackage{axodraw}
\usepackage{graphics}
\usepackage{subfigure}
\usepackage{epstopdf}

\newcommand\squark{{\tilde q}}
\newcommand\gluino{{\tilde g}}

\newcommand{\pt}{\ensuremath{p_{T}}}
\newcommand{\GeV}{\ensuremath{\text{GeV}}}
\newcommand{\TeV}{\ensuremath{\text{TeV}}}
\newcommand{\ipb}{\ensuremath{\text{pb}^{-1}}}
\newcommand{\ifb}{\ensuremath{\text{fb}^{-1}}}

\newcommand{\mzero}{\ensuremath{m_{0}}}
\newcommand{\mhalf}{\ensuremath{m_{1/2}}}
\newcommand{\mth}{\ensuremath{m_{3/2}}}

\graphicspath{{plots/}}

\author[a]{B.C.~Allanach,}
\affiliation[a]{Department of Applied Mathematics and Theoretical Physics, Centre for Mathematical Sciences, University of Cambridge, Wilberforce Road, Cambridge CB3 0WA, United Kingdom}
\emailAdd{B.C.Allanach@damtp.cam.ac.uk}

\author[b]{T.J.~Khoo,}
\affiliation[b]{Department of Physics,
Cavendish Laboratory,
J J Thomson Avenue,
Cambridge,
CB3 0HE, United Kingdom}
\emailAdd{khoo@hep.phy.cam.ac.uk}

\author[a,b,c]{K.~Sakurai}
\affiliation[c]{Department of Physics,
Nagoya University,
Nagoya 464-8602, Japan}
\emailAdd{sakurai@hep.phy.cam.ac.uk}

\title{Interpreting a 1 fb$^{-1}$ ATLAS Search in the Minimal Anomaly
  Mediated Supersymmetry Breaking Model}

\abstract{Recent LHC data significantly extend the exclusion limits for
  supersymmetric particles, particularly in the jets plus missing transverse
  momentum channels. The most recent such data have so far been interpreted by
  the   experiment in only
  two different supersymmetry breaking models: the 
  constrained minimal supersymmetric standard model (CMSSM) and a simplified 
  model with only squarks and gluinos and massless neutralinos. We compare
  kinematical distributions of supersymmetric signal events predicted by the
  CMSSM and anomaly mediated supersymmetry breaking (mAMSB) before
  calculating exclusion limits in mAMSB\@.
  We obtain a lower limit of 
  900 GeV on squark and gluino masses at the 95$\%$ confidence level
  for the equal mass limit,  $\tan \beta=10$ and $\mu>0$. 
}

\begin{document}
\maketitle

\section{Introduction}

Initial supersymmetric particle search data from ATLAS
\cite{ATLAS:2011qk,ATLAS:2011hh} and CMS \cite{Khachatryan:2011tk} 
have now been extended from 
35~\ipb{} to 1~\ifb{}~\cite{newATLAS,alphaTCMS, mt2CMS, mhtCMS}.
The data, collected from $pp$ collisions at $\sqrt{s}=7$ TeV,
feature significant missing transverse momentum and jet activity.
Other topologies have also been examined, but in most scenarios of supersymmetry
breaking, the jets plus missing transverse momentum signatures are the most sensitive.
None of these searches has found a significant signal over the expected
Standard Model (SM) background, and so they have set limits on
sparticle production.
Within the CMSSM \citemsugraandcmssm{}, the strongest limits come from the 
ATLAS ``0-lepton'' search \cite{newATLAS} which excludes 
squarks and gluinos with masses below 950 GeV (in the equal mass limit) at
95$\%$ C.L. in the $A_0=0$, $\tan(\beta)=10$, $\mu>0$ slice of the CMSSM\@,
and the CMS hadronic searches that extend the mass limit to 
1.1~\TeV{} ~\cite{alphaTCMS, mt2CMS, mhtCMS}.  A limit at this scale was
already anticipated
in the absence of a signal~\cite{Bechtle:2011dm}.

The experiments present their CMSSM limits on the
scalar-gaugino universal mass $(m_0,\ m_{1/2})$ plane for
particular values of $\tan \beta$ (the ratio of the two MSSM Higgs vacuum
expectation values) and $A_0$ (a GUT-scale universal tri-linear scalar
coupling parameter). 
Earlier LHC supersymmetry (SUSY) searches based on only 35~\ipb of integrated luminosity
have been reinterpreted for different values of $A_0$ and $\tan \beta$ from
those presented by the experiments~\cite{Akula:2011zq}. 
The earlier searches were also reinterpreted
in terms of gauge mediation and some benchmark SUSY
breaking models~\cite{Dolan:2011ie}. Searches based on 165~\ipb{} have also
been investigated in a SUSY breaking model which has certain
non-universalities motivated by naturalness in electroweak symmetry
breaking~\cite{Sakurai:2011pt}. The first \ifb{} of preliminary LHC data has
been 
used to confront the no-scale $F-SU(5)$ model \cite{Nano:2011uf} and, more
recently, the data were published and exclusion limits in the
phenomenological MSSM~\cite{Sekmen:2011cz} were calculated, combining 
channels with leptons, jets and missing transverse momentum with the jets plus
missing transverse momentum channels.
The present paper is in a similar spirit to these earlier works:
it is our aim to assess the impact of the most recent published 0-lepton 
results based on 1~\ifb{} of integrated luminosity on the mAMSB
model~\cite{Randall:1998uk}. 
We also wish to study the effectiveness of the 
experimental cuts in the context of mAMSB\@.
mAMSB is a model worthy of study,
since it avoids the flavour and SUSY CP
problems~\cite{Gherghetta:1999sw,Feng:1999hg,Allanach:2009ne} by introducing the 
additional universal soft mass $m_0$ and can provide
the correct relic density of dark 
matter in a natural way in a sequestered hidden sector~\cite{Feng:2011ik,Feng:2011uf}.

ATLAS also presented their exclusion limits in the squark-gluino mass
plane~\cite{newATLAS}, 
assuming all other sparticles are heavy, except for the the neutralino which
is massless. One 
may ask whether one should just take this simplified model, and apply it to
the parameter space of mAMSB. Each point in mAMSB parameter space corresponds
to some gluino and squark mass and a light neutralino and so, in principle,
one could just chart the approximate exclusion in the mAMSB parameter space
without performing any event simulation, instead only calculating the sparticle
spectrum. We shall show that this 
leads to a poor approximation of the 0-lepton search exclusion
for the full mAMSB model. Thus, we shall need to simulate sparticle production
in mAMSB. 

Early work on collider signatures of mAMSB focused on the 
usual supersymmetric signatures~\cite{Paige:1999ui} as well as 
searching for the decays
of the lightest charginos into the lightest neutralinos~\cite{Barr:2002ex}. 
These
two particles being close in mass is a prediction of mAMSB, and the hope was
to confirm this by measuring the decays of one into the other. 
The LHC
experiments have so far not yet published results using the special techniques of searching for the
lightest chargino decays, and so we do not comment on them further. 
It was also recently shown that a light CMSSM point containing the
``golden'' supersymmetric decay cascade
chain $\tilde q \rightarrow \chi_2^0 \rightarrow {\tilde e}_R \rightarrow \chi_1^0$
could be distinguished against mAMSB on the basis of kinematic end-point
measurements alone~\cite{Allanach:2011ya} with just 10 \ifb{} of LHC
data. Unfortunately, such a light CMSSM point has been excluded by the 2011
LHC searches and no sign of the golden chain has been seen.

The paper proceeds as follows: we present the cuts and summarise the results
of the ATLAS 0-lepton search in section~\ref{sec:0lep}. Next, in
section~\ref{sec:simulation}, we describe our
simulation of the ATLAS search, first validating our simulation against their
results in the CMSSM\@. We then simulate mAMSB signals, putting them through the
same analysis in section~\ref{sec:amsb}. We investigate the 
properties of interest of mAMSB signal events and compare them to those of the
CMSSM\@. We then calculate the main result of our paper: the exclusion limit in
mAMSB\@. The comparison with the simplified model approximation is performed,
finding it to be a poor approximation to the full mAMSB exclusion.
Finally, in section~\ref{sec:summ}, we summarise and conclude the
paper. 

\section{The ATLAS 0-lepton Search\label{sec:0lep}}

\begin{table}
\begin{center}
\begin{tabular}{l|lllll}
\hline
 & $\geq 2$ jets & $\geq 3$ jets& $\geq 4$ jets& $\geq 4$ jets'& High mass \\
\hline
 $p_{T}(j_1)$ & $>130$ GeV & $>130$ GeV & $>130$ GeV & $>130$ GeV& $>130$ GeV\\
$p_T(j_2)$ & $>40$ GeV & $>40$ GeV & $>40$ GeV & $>40$ GeV& $>80$ GeV\\
 $p_T(j_3)$ & $-$  & $>40$ GeV & $>40$ GeV & $>40$ GeV& $>80$ GeV\\
$p_T(j_4)$ & $-$  & $-$  & $>40$ GeV & $>40$ GeV& $>80$ GeV\\
 $|\ptmiss|$ & $>130$ GeV& $>130$ GeV& $>130$ GeV& $>130$ GeV& $>130$ GeV\\
 $\Delta\phi$ & $>0.4$& $>0.4$& $>0.4$& $>0.4$& $>0.4$ \\
 \ptmiss{}$/\meff$ & $>0.3$ & $>0.25$& $>0.25$ & $>0.25$ & $>0.2$\\
 $\meff$ & $>1000$ GeV & $>1000$ GeV & $>500$ GeV& $>1000$ GeV& $>1100$ GeV\\
\hline
Observed & 58 & 59& 1118& 40 & 18\\
Background & 62.4$\pm$4.4$\pm$9.3 & 54.9$\pm$3.9$\pm$7.1 &
1015$\pm$41$\pm$144 & 33.9$\pm$2.9$\pm$6.2 & 13.1$\pm$1.9$\pm2.5$ \\
\hline
$\sigma \times A \times \epsilon $/fb & 22 & 25 & 429 & 27 & 17 \\
\hline
\end{tabular}
\end{center}
\caption{The cuts used to define the signal regions of the ATLAS-0-lep
  analysis \cite{newATLAS}.  $\Delta \phi$ is the minimum azimuthal angle
  between $\ptmiss$ and the first three jet $p_T$s. 
  The jet $p_T$ thresholds, $p_T(j_n)$ are ordered in decreasing order.
  We also display the number of events ATLAS observed in each region,
  along with the expected Standard Model backgrounds. The first uncertainty
  represents the statistical uncertainty on the background, whereas the second
  labels the 
  systematic uncertainty. In the final row, we show the 95\% C.L. exclusion upper
  limit on the total SUSY cross-section times acceptance times efficiency for
  each signal   region, from Ref.~\protect\cite{newATLAS}.
\label{tab:atlascuts}}
\end{table}

The ATLAS collaboration based its most recent search on {\em five}\/ sets of cuts
on {\em two}\/ different variables (the effective mass, $\meff$,
\cite{Hinchliffe:1996iu,Tovey:2000wk} and the magnitude of the missing
transverse momentum $|\ptmiss|$) which have 
properties tailored more specifically to the kinematic properties of
$\squark\squark$, $\squark\gluino$ and $\gluino\gluino$ production.
$\meff$ is defined as the sum of $|\ptmiss|$ and the magnitudes of the
transverse momentum of the two, three or four highest $p_T$ jets, depending on
whether the signal region specifies greater than two, three or four jets,
respectively.  Alternatively, an \emph{inclusive} $\meff$ variant summing $|\ptmiss|$
and the $p_T$ of all jets with $p_T > 40~\GeV$ is defined for the ``high mass'' selection
requiring four jets with a tighter $p_T$ threshold of 80~\GeV.

The cuts defining the search regions used by the ATLAS 0-lepton analysis are
given in Tab.~\ref{tab:atlascuts}. 
Also shown for each signal region are the number of observed events $n_o^{(i)}$
that made it past 
cuts and the expected SM backgrounds $n_b^{(i)}$ together
with their statistical and systematic errors $\sigma_{b,~\text{stat}}^{(i)}$
and $\sigma_{b,~\text{syst}}^{(i)}$.
$\sigma_{b,~\text{syst}}^{(i)}$ incorporate various experimental and theoretical
uncertainties on the background predictions, notably those due to the
jet energy scale and resolution, and uncertainties on Monte Carlo modelling.

ATLAS constructed frequentist exclusion regions in SUSY parameter space using
a profile  
likelihood ratio method, taking into account theoretical and detector systematics. The
information from the five signal regions was 
combined by defining the test statistic of each parameter point to be a
likelihood ratio 
given by the signal region demonstrating the best expected sensitivity to new
physics. 
Results were presented as 95\% Confidence Level (C.L.) exclusion regions in the
$(m_{\tilde{g}},m_{\tilde{q}})$ plane for $m_{\chi_1^0}=0$ and in the $\tan
\beta=10$, $A_0=0$, $\mu>0$ slice of the CMSSM~\cite{newATLAS}. 

\section{Our Simulation of the ATLAS Search \label{sec:simulation}}

In order to estimate SUSY exclusions, we must simulate LHC collisions
producing SUSY particles, along with their subsequent decays. In order to do
this, 
the sparticle spectrum is produced with {\tt
  SOFTSUSY3.1.7}~\cite{Allanach:2001kg} in SUSY Les Houches Accord
format~\cite{Skands:2003cj}.  
We simulate the production and decay of sparticles using {\tt
  HERWIG++2.5.1}~\cite{Bahr:2008pv}, and detector simulation is by {\tt
  DELPHES1.9}~\cite{Ovyn:2009tx} using a modified ATLAS detector card. 
Simulating only squark and gluino production for the CMSSM is sufficient,
since direct production of neutralinos, charginos or sleptons is negligible.
However, in mAMSB, the direct associated production of squarks/gluinos with
charginos or neutralinos is significant, and we include these channels in our
mAMSB simulation.
Jets are defined using the anti$-k_T$ algorithm with $\Delta R=0.4$ and an
energy recombination scheme in {\tt
  FASTJET2.4.3}~\cite{fastjet,Cacciari:2005hq}.  
The total SUSY production cross-sections are computed at next-to-leading order in
{\tt PROSPINO 2.1}~\cite{prospino}.
For each point in parameter space, we simulate 10000 SUSY events in order to
deduce expectations for signal yields and kinematic properties. We shall see
below that the efficiencies we obtain are typically higher than 10$\%$ in each
different signal region, meaning that statistical fluctuations in our expected
signal yield are negligible.

The ATLAS analysis cuts detailed in Tab.~\ref{tab:atlascuts} do not include the 
details of a correction for data corrupted by the loss of some calorimeter
regions  in ATLAS during part of the data-taking period.
To approximate the impact of this additional event cleaning, we apply an
estimated correction factor of $A=0.85$ to the signal acceptance.

As ATLAS computes the limits on the CMSSM parameter space by using a
sophisticated 
likelihood function to run hypothesis tests, it is unfeasible to reinvent
their statistical 
methods without access to the finer details of their signal and
backgrounds. Instead, 
we estimate the exclusion reach of the selection by comparing with ATLAS' quoted
model-independent limits on $(\sigma\times A \times \epsilon)$ for each signal selection
shown in Table~\ref{tab:atlascuts}.
A model point is considered to be excluded if the value we compute for 
$(\sigma\times A\times\epsilon)$ exceeds the limits produced by ATLAS\@.
In this way, we take into account the full power of ATLAS' statistical
infrastructure, 
except for the impact of signal systematics on the limit, 
which we model as specified below.

The assumptions and approximations we have made in reproducing ATLAS' analysis
are tested as detailed in the next section.

\subsection{Validation of Our Simulation \label{sec:val}}

\begin{figure}[h!]\begin{center}

\subfigure[2 jets]{\includegraphics[width=0.4\textwidth]{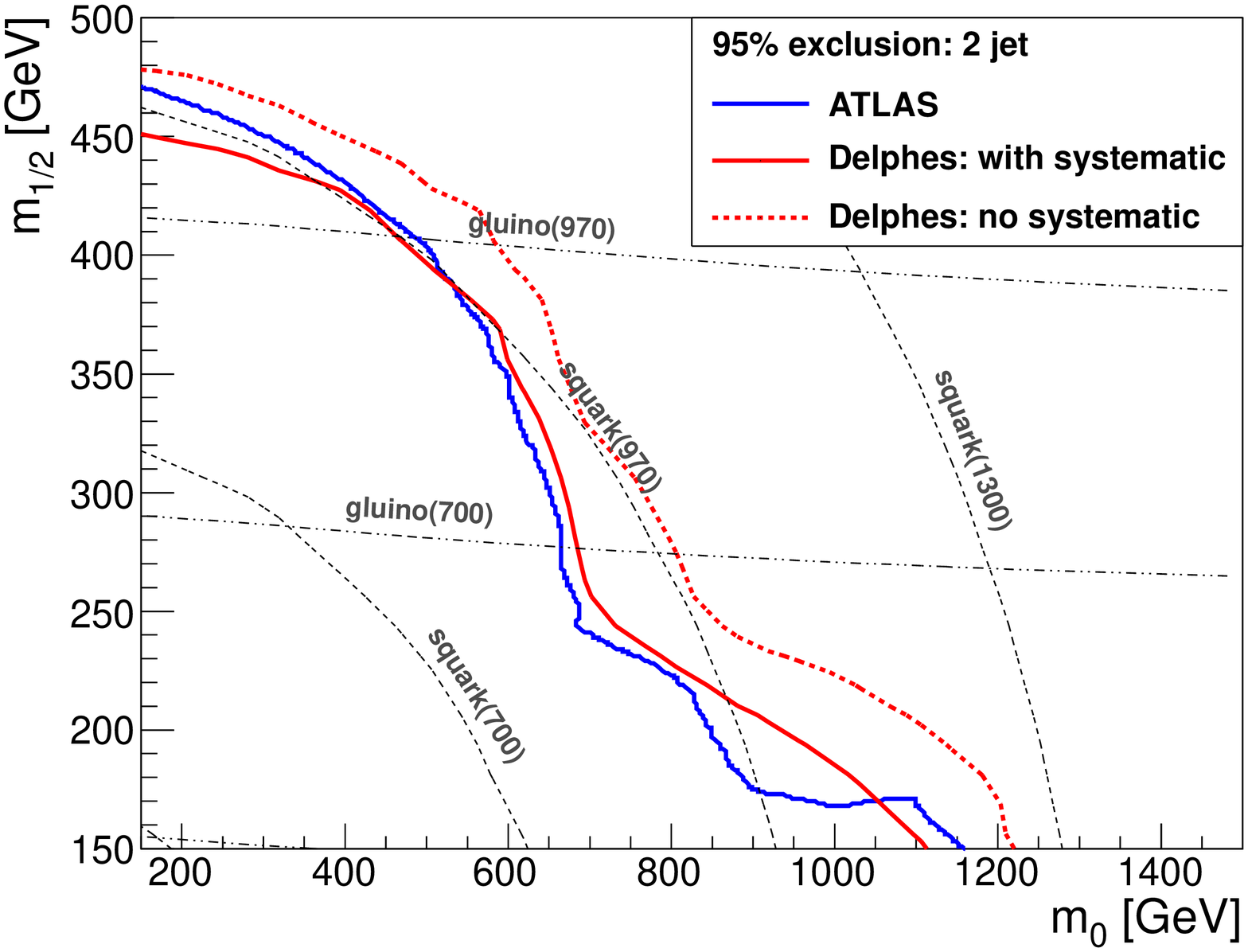}}
\hspace{2ex}
\subfigure[3 jets]{\includegraphics[width=0.4\textwidth]{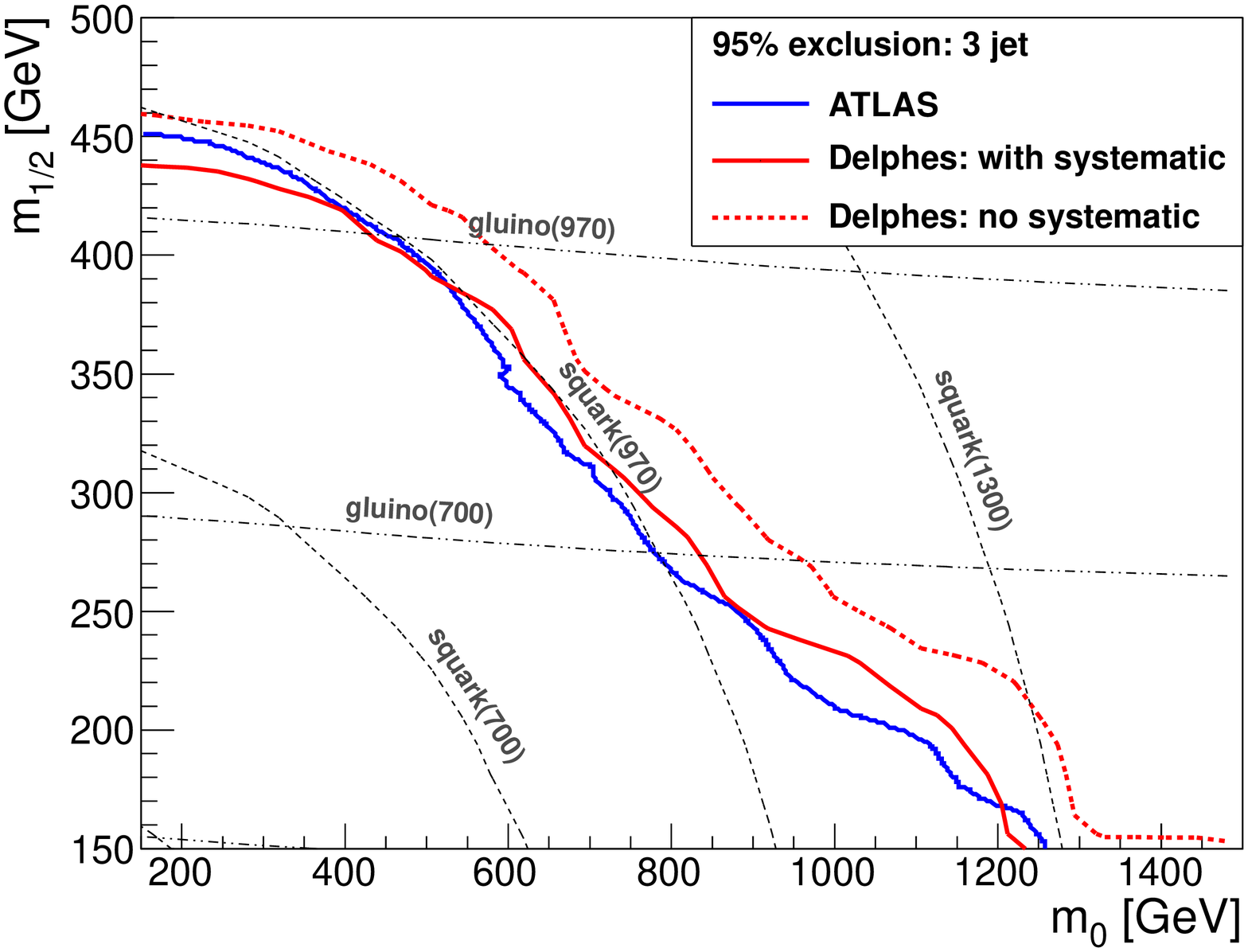}}

\subfigure[4 jets]{\includegraphics[width=0.4\textwidth]{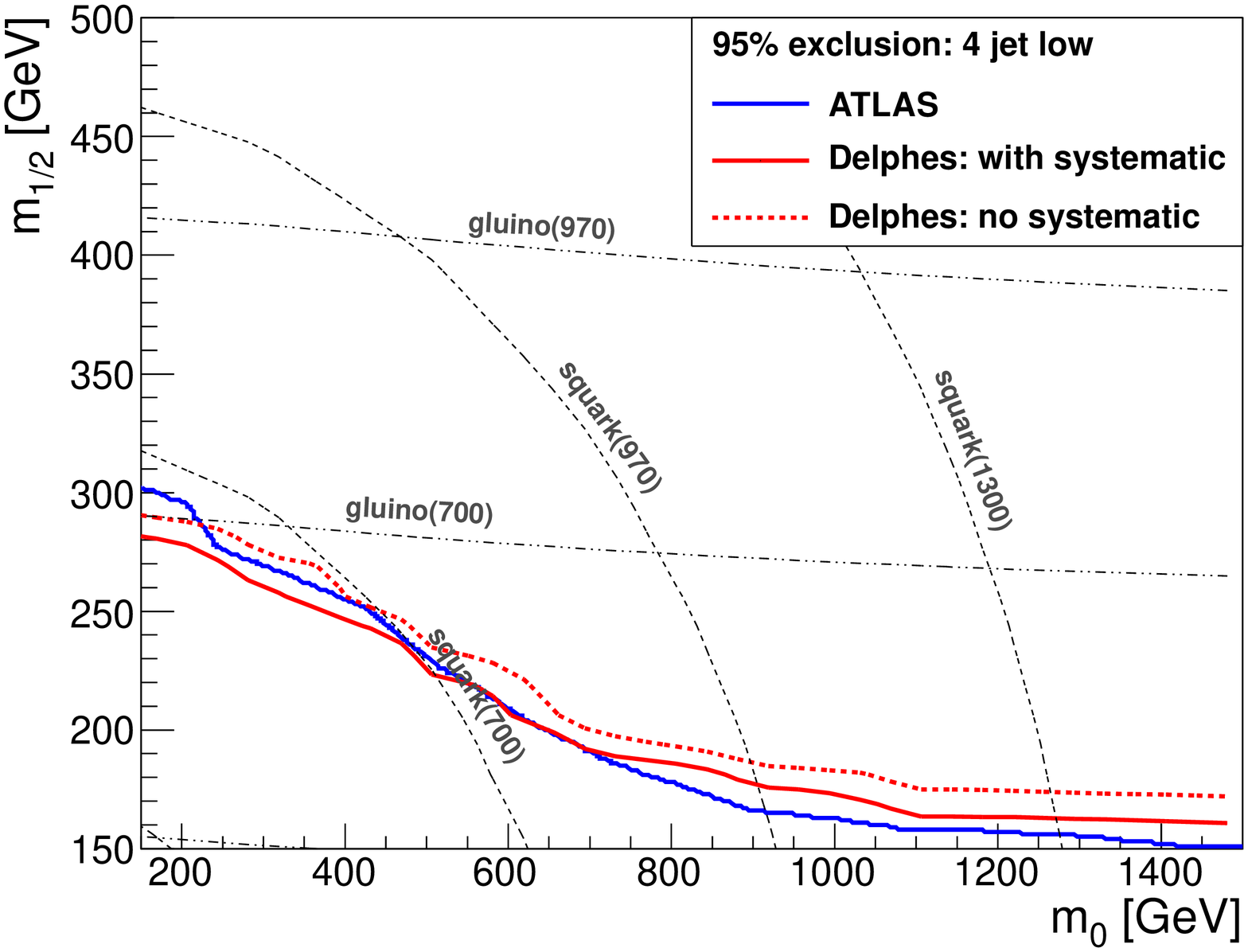}}
\hspace{2ex}
\subfigure[4 jets']{\includegraphics[width=0.4\textwidth]{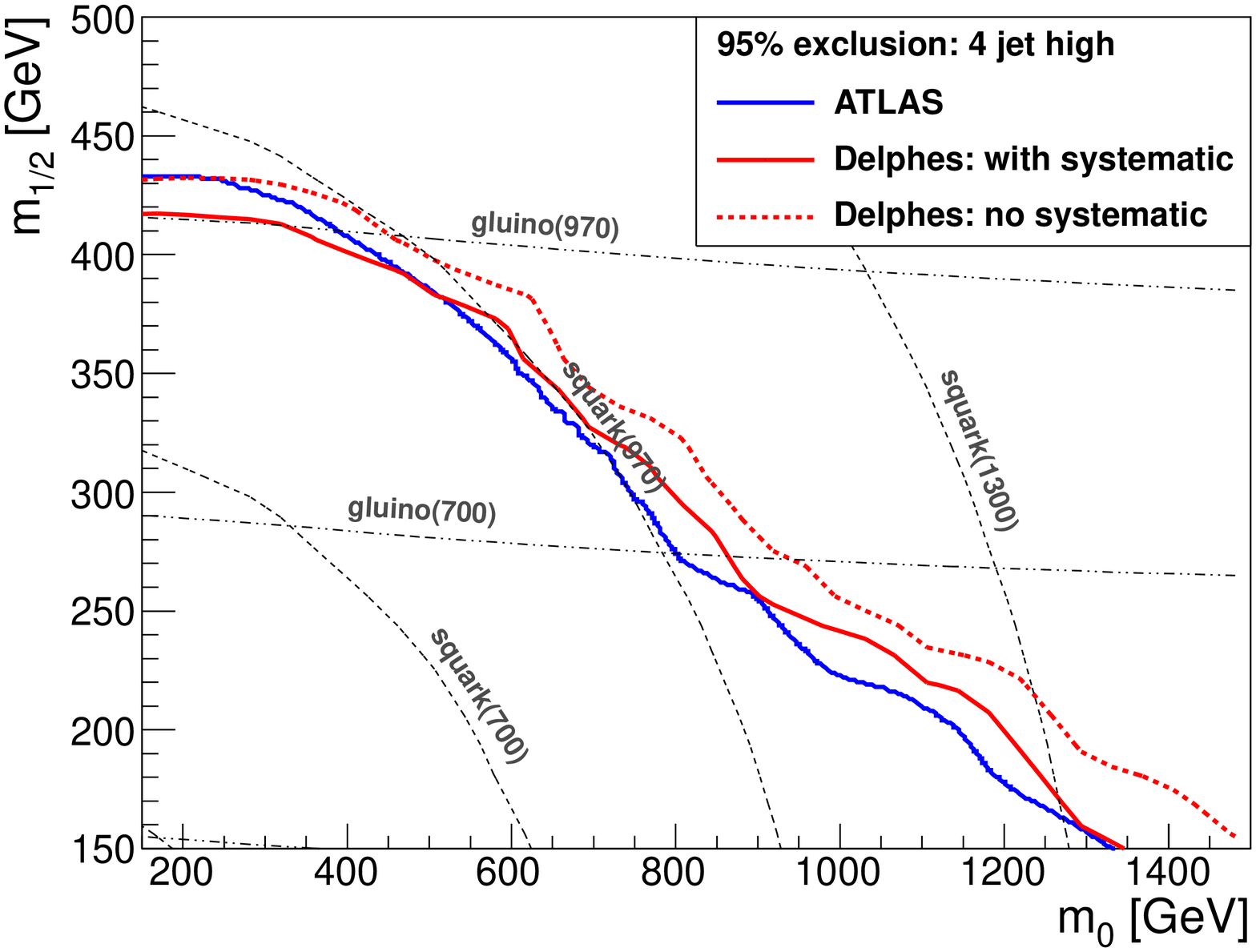}}

\subfigure[high mass]{\includegraphics[width=0.4\textwidth]{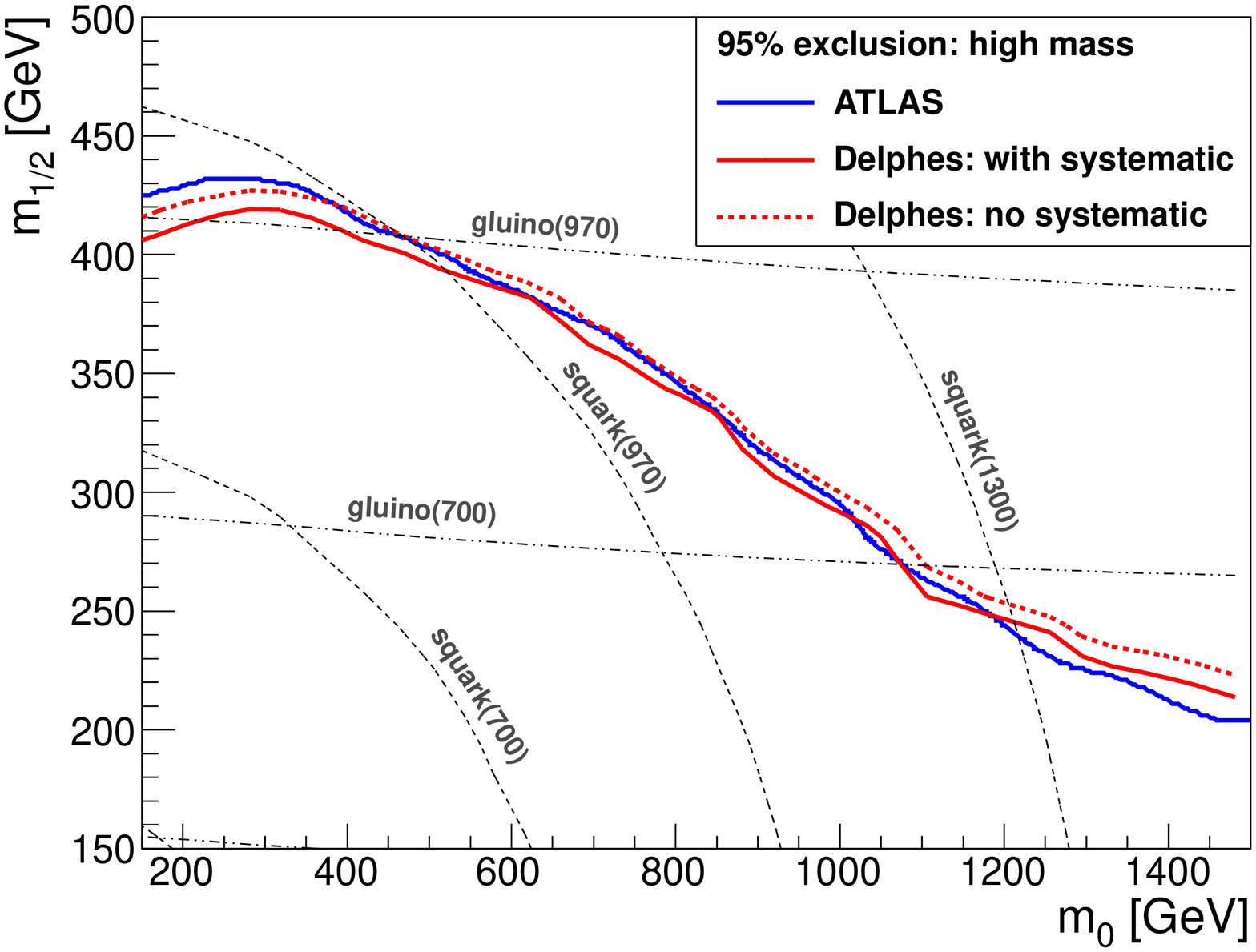}}

\caption{\label{fig:val} Comparison of our $95\%$ C.L. exclusion limits with those of 
  ATLAS in the case of the CMSSM with $\tan \beta=10$, $A_0=0$ and $\mu>0$. Each
  sub-figure shows a different signal region, as defined in
  Table~\protect\ref{tab:atlascuts}. One solid curve shows our estimate
  including signal systematic errors,
  the other shows ATLAS'~\cite{newATLAS}.
  The dashed curve shows our estimate neglecting systematic errors in the
  signal.   We
  show iso-contours of gluino and squark mass as labelled dotted lines.
} 
\end{center}\end{figure}

We cannot match the accuracy of ATLAS' exclusion in parameter space, because
we do not have access to a detailed detector simulation. We do perform a more 
approximate detector simulation with {\tt
  DELPHES1.9}, but it is important for us to validate this approximation 
in order to find out how good it is. Importantly, we do not calculate signal
systematic errors explicitly in our initial determination: this is impossible
for us to 
do because we are using the quoted ATLAS bounds on $\sigma \times \epsilon
\times A$, which also do not include such systematic
errors\footnote{Systematic errors upon the background are included,
  however.}. However, in the exclusion contours, ATLAS does determine and
include the signal systematic. 
The signal 
systematic errors come from uncertainties in the 
parton density and from higher order corrections and from uncertainties 
due to initial state radiation modelling and other jet modelling and
measurement effects. 
If we had enough information to reconstruct the likelihood, we
could make an attempt to calculate some components of the systematic error. 
Instead, we here perform a rough `by eye' fit, allowing a different
systematic error 
for each signal region that does not depend on the SUSY breaking parameters.
The systematic error in the likelihood changes $\epsilon$ by a factor of $s_R$
for each signal 
region $R$.
We then estimate a measured cross-section of $\sigma \times \epsilon \times A
\times s_R$ for each signal region $R$. 

We now present the validation of our determination of the exclusion by
comparing our approximation with that of ATLAS' in the case of the CMSSM\@. 
ATLAS determined the exclusion for $\tan \beta=10$, $A_0=0$ and $\mu>0$ in the
CMSSM~\cite{newATLAS}. We use ATLAS' quoted $95\%$ C.L. bounds on $\sigma \times
A \times \epsilon$, where we calculate $\epsilon$ from our simulation and
$\sigma$ from {\tt PROSPINO 2.1}~\cite{prospino},
displaying the contour as a solid line in Fig.~\ref{fig:val}, where we have
included the systematic error in each signal region as stated above. 
We find that signal systematic error factors of $s_{2j}=s_{3j}=0.7$,
$s_{4j}=s_{4j'}=0.8$ and $s_{hm}=0.9$ reproduce the ATLAS exclusion contours
quite well, as the 
figure shows. Our determination of the exclusion contours {\em neglecting} 
systematic error is shown as the dashed contours. These clearly are a
worse approximation.
In the equal squark-gluino mass limit, and using our approximate exclusion
limits, we obtain a lower bound on the mass of 970 GeV at 95$\%$ CL from
Fig.~\ref{fig:val}a\footnote{If we were to use the zero
signal-systematic contours, we would obtain a badly determined bound of
1020~GeV.}. This combined exclusion limit is defined by using the most
restrictive signal region at the parameter point in question. For the equal
squark-gluino mass limit, the most sensitive region is the 2 jets region.
We see that at this point, the ATLAS exclusion lays on top of our exclusion.
However, ATLAS quotes an equivalent 
bound of 950~GeV, close but different to our determination. This 20 GeV
difference is due to the different SUSY spectrum generators used: whereas {\tt
  SOFTSUSY3.1.7} is used here, ATLAS used {\tt
  ISASUSY7.80}~\cite{Paige:2003mg}. Such differences are 
caused by higher order corrections in the respective calculations, and as such
form part of the theoretical error~\cite{Allanach2003jw}. 
By examining the difference between the ATLAS exclusion limits and our
estimate, we estimate an error in our exclusion of around  
30 GeV in the squark
and gluino masses.  
We see that the signal regions which are most sensitive are the 2-jet search
region at high $m_{1/2}$ and low $m_0$, and the high mass region at large
$m_0$ and low $m_{1/2}$.
In each signal region, our estimate of the exclusion is similar to that of
ATLAS, 
and we conclude that our approximation is reasonable. We should
therefore be able to re-simulate signal events in different supersymmetry
breaking scenarios in order to evaluate  exclusion limits upon them. We now 
perform this task in mAMSB\@. 

\section{mAMSB Simulation\label{sec:amsb}}

We now simulate the SUSY signals for a grid of mAMSB points for $\tan
\beta=10$ and $\mu>0$ in mAMSB, applying the ATLAS cuts to them as we did for
the CMSSM, above and using the signal systematic errors as determined in our
validation, section~\ref{sec:val}. In mAMSB, we scan over $m_0$ and $\mth$, the
auxiliary mass, 
which the AMSB soft SUSY breaking terms are proportional to (for further
details on the connection between these parameters and the SUSY spectrum, see
Ref.~\cite{Gherghetta:1999sw}). In practice, limited mainly by disk storage
(since the event files are very large), we scan in an 11 by 11 grid. 

To get a sense of the mAMSB model characteristics, we plot some relevant
mass parameters across the parameter plane in Fig.~\ref{fig:AMSBmasses}.
The gluino and squark mass variation is similar to that shown by the CMSSM\@.
Between the chargino next-to-lightest supersymmetric particle (NLSP) and
neutralino lightest supersymmetric particle (LSP), there is only a small mass
gap of a few hundred MeV, which is also illustrated in
Fig.~\ref{fig:AMSBmasses}. 

The SUSY signal generation is restricted to processes producing at least one
strongly-interacting sparticle. This is to ensure that the events are not
dominated 
by the production of two LSPs or NLSPs, as such events are effectively
invisible. 
As long as the squark and gluino masses are relatively low, i.e. below a TeV,
strong production dominates the remaining SUSY cross-section. Above this mass
scale, the production shifts towards electroweak associated production of one
gaugino and one squark or gluino. This potentially impacts on the efficiency
of the signal selection as the mass scale grows, although heavy sparticles are
typically quite visible due to the fixed and large mass-splittings.

The largest direct production total SUSY cross-section at the mAMSB point is
$\tilde g \chi_1^\pm$ production at 11 fb, but many other processes are
important, for instance $\tilde g  {\tilde u}_L$ production at 10 fb. 
At the CMSSM point, the largest SUSY direct production cross-sections are 
$\tilde g  {\tilde u}_R$ production at 9 fb, $\tilde g {\tilde u}_L$
production at 7 fb and ${\tilde u}_L {\tilde d}_L$ production at 6 fb. 
Direct squark/gluino production in association with a weak-ino is important
at the mAMSB point (in contrast with the CMSSM). 
Fig.~\ref{fig:AMSBewfrac} illustrates the variation of the fraction of electroweak associated
production over the total SUSY cross-section generated. At large \mzero{} and large
\mth{}, the cross-section is entirely due to associated production.

\begin{figure}[htbp!]
\begin{center}

\subfigure[Average light flavour squark mass]{\includegraphics[width=0.45\textwidth]{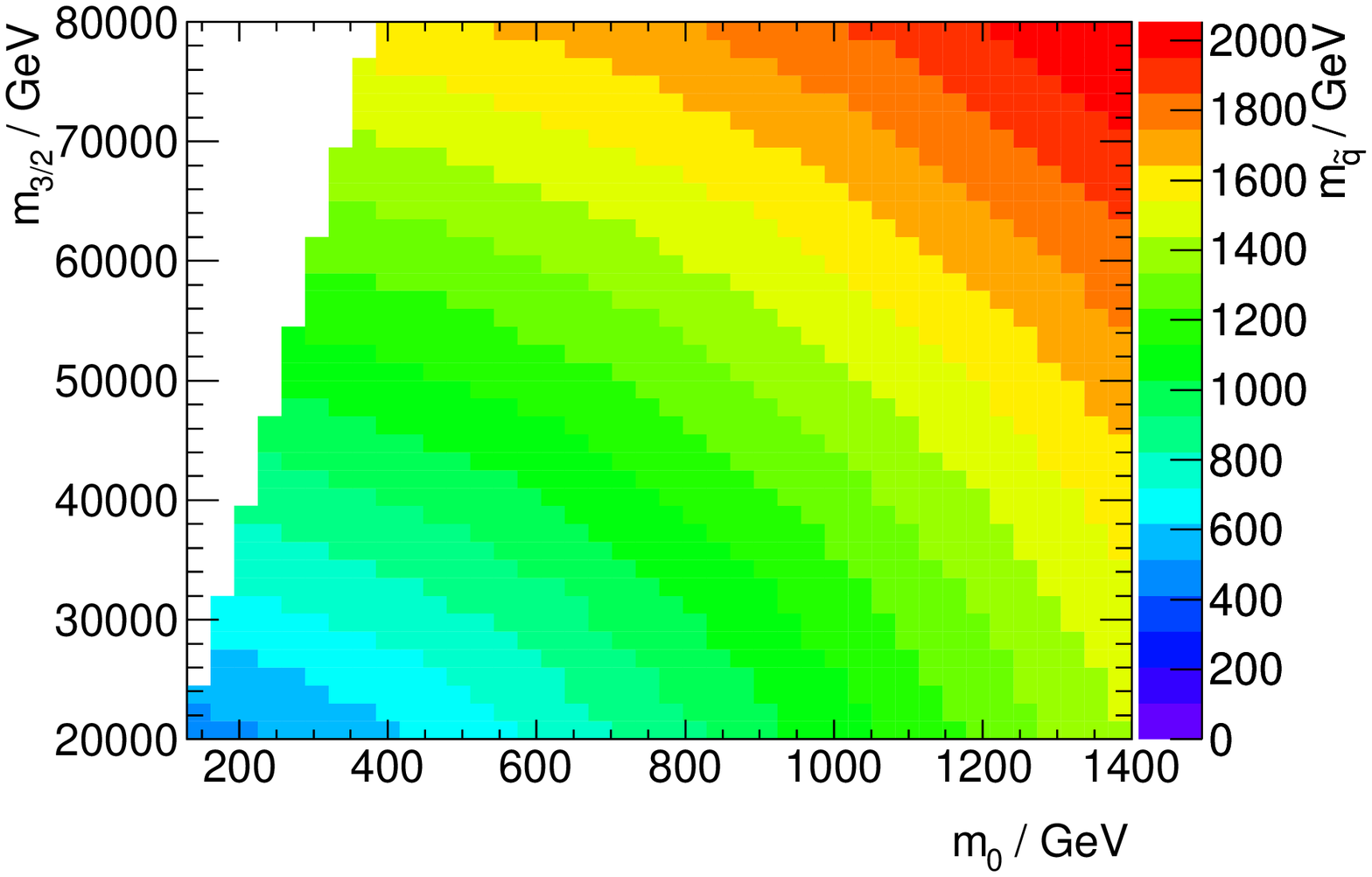}}
\hspace{2ex}
\subfigure[Gluino mass]{\includegraphics[width=0.45\textwidth]{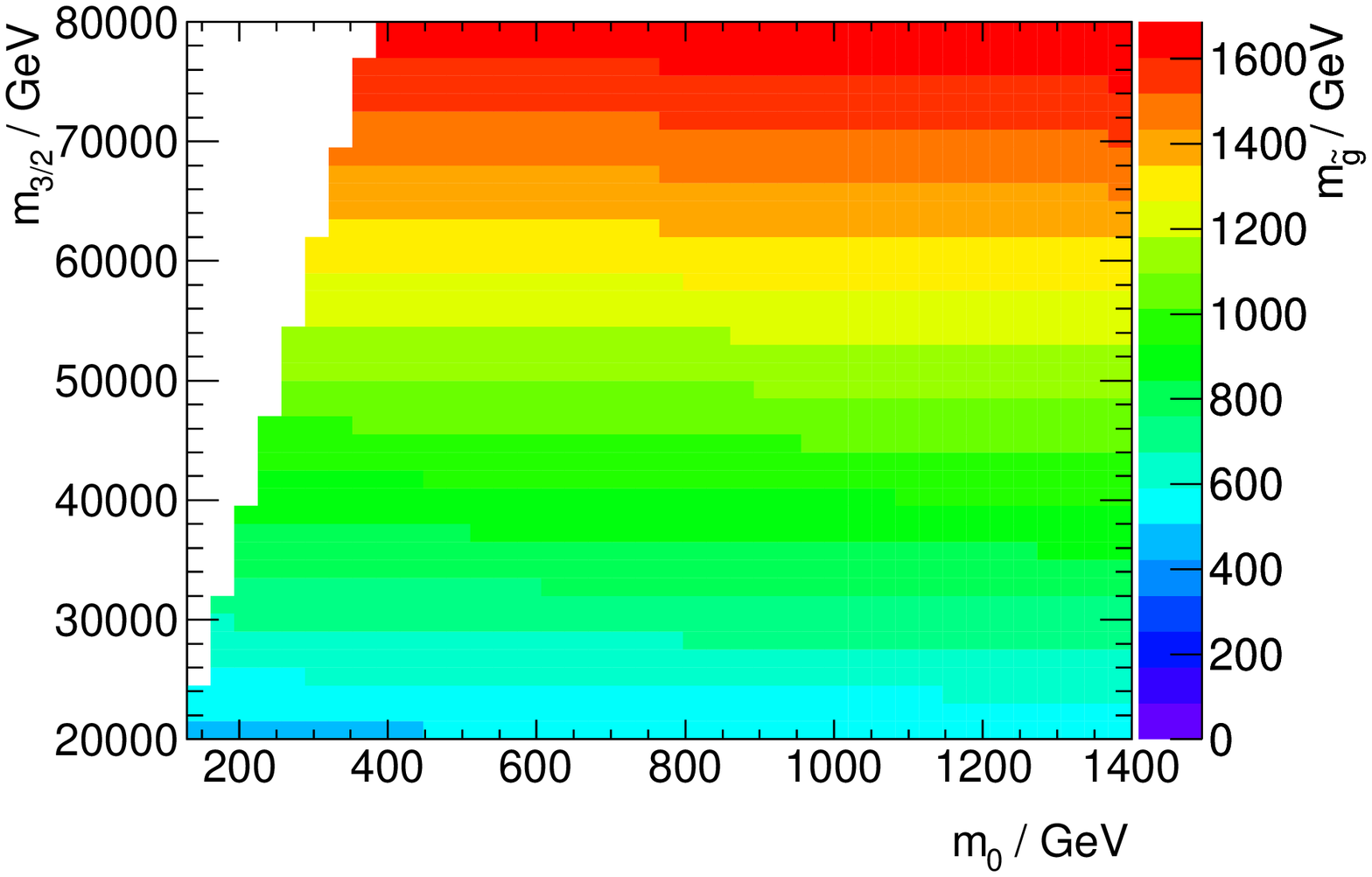}}

\subfigure[LSP mass]{\includegraphics[width=0.45\textwidth]{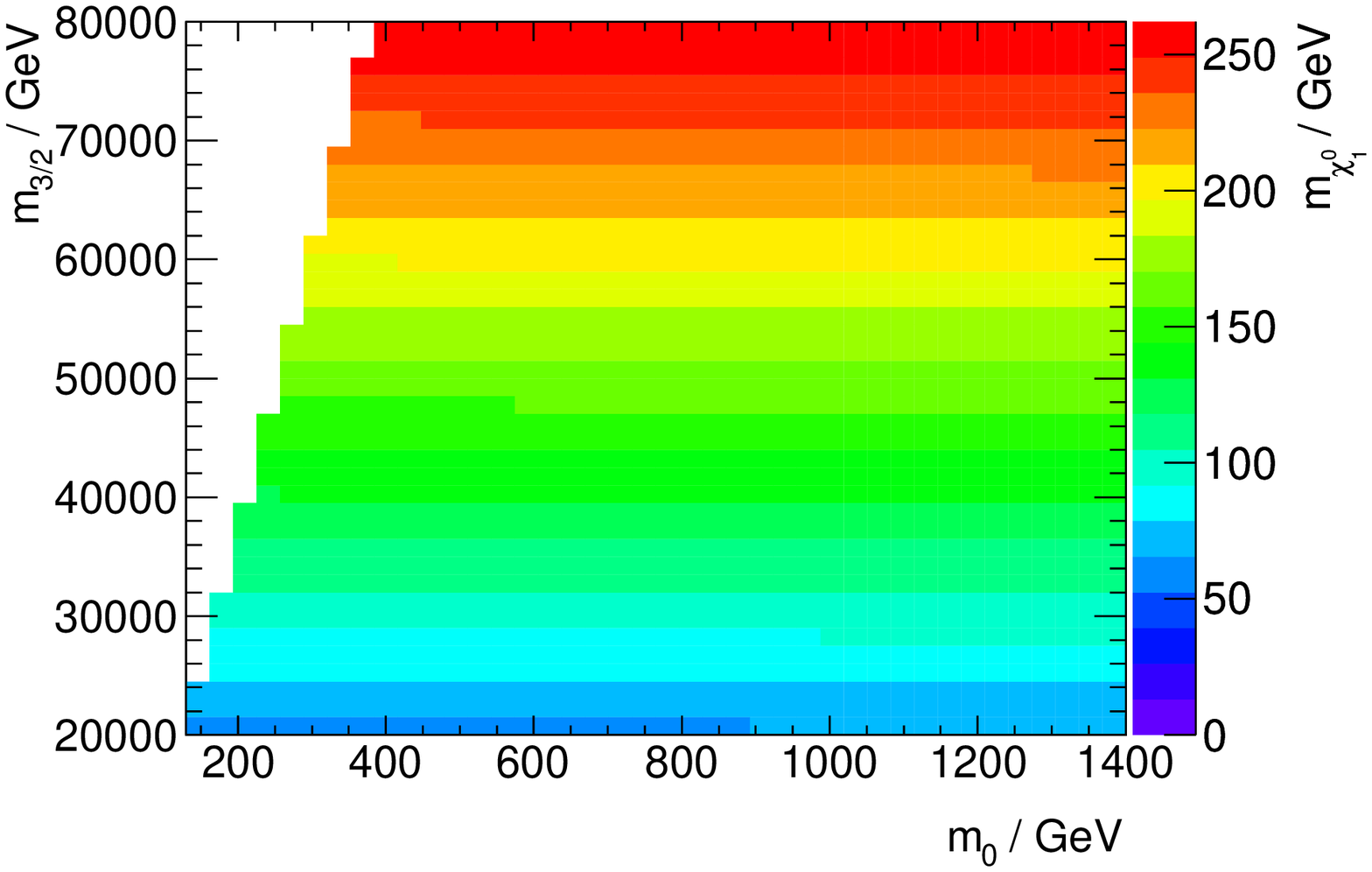}}
\hspace{2ex}
\subfigure[LSP-NSLP mass difference]{\includegraphics[width=0.45\textwidth]{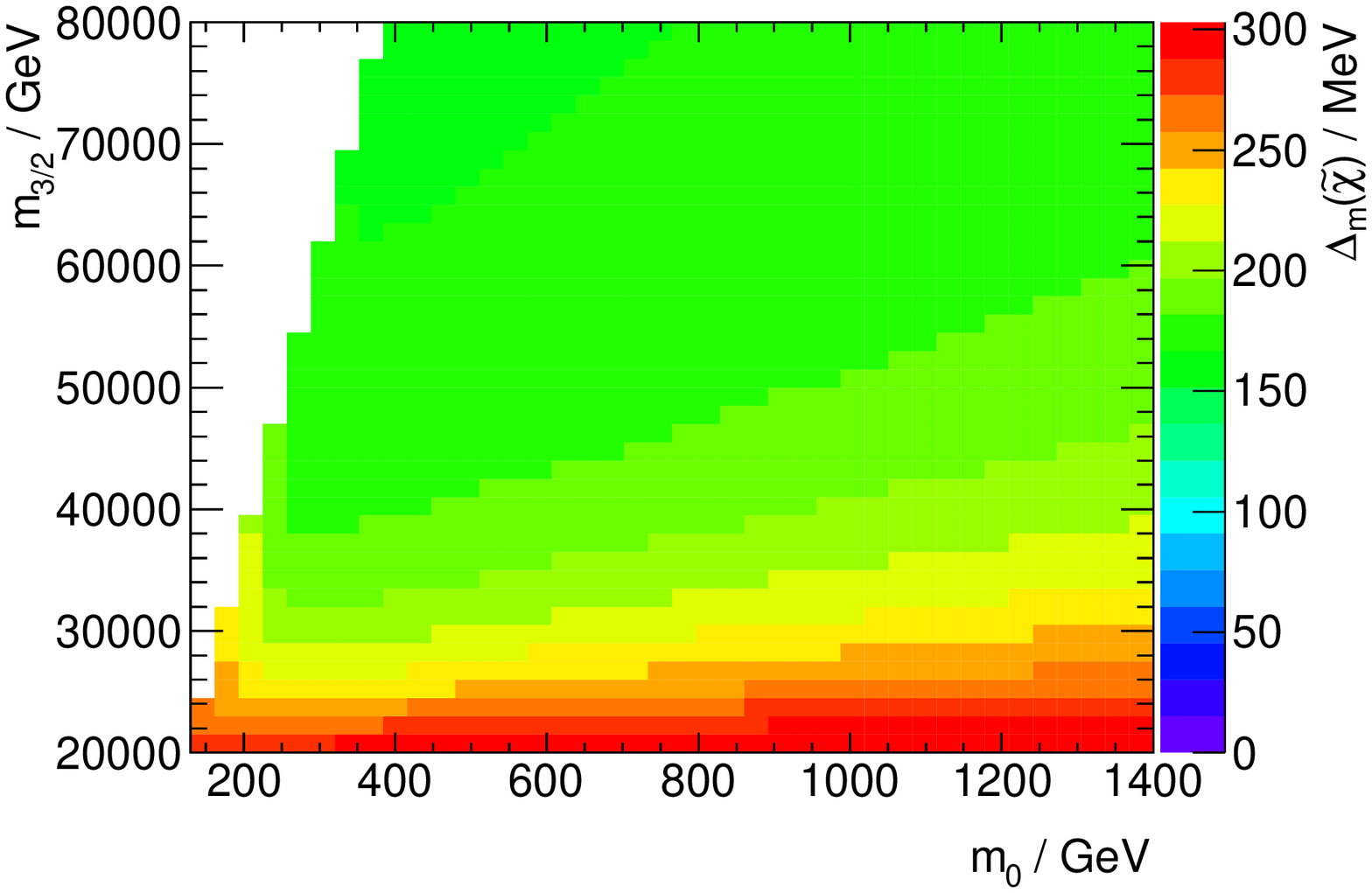}}

\caption{Masses of the squarks, gluinos and LSP (a-c) and the mass-splitting
  between the LSP and chargino NLSP (d) across the mAMSB $\mzero-\mth$ plane
  for $\tan \beta=10$ and $\mu>0$.
The white region in the top left of each plot is theoretically excluded due
to the presence of negative mass-squared scalars. }
\label{fig:AMSBmasses}
\end{center}
\end{figure}

\begin{figure}[htbp!]
\begin{center}

\includegraphics[width=0.45\textwidth, clip, trim = 0 0 0 1.25cm]{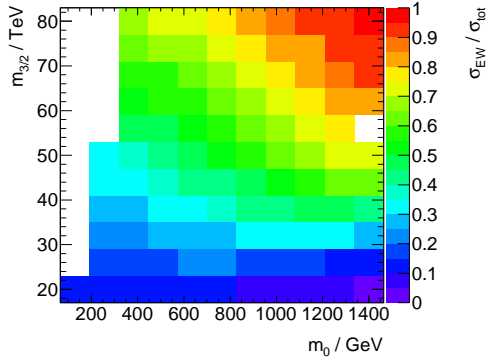}

\caption{Variation across the mAMSB $\mzero-\mth$ plane of the fraction of
  electroweak production over the total SUSY production cross-section,
  requiring at least one squark or gluino to be produced. The white region in 
  the top left of each plot is theoretically excluded due to the presence of negative 
  mass-squared scalars, while the white spot on the right is due to a failed MC job.}
\label{fig:AMSBewfrac}
\end{center}
\end{figure}

\subsection{Comparison of Signal Events Between the CMSSM and mAMSB \label{sec:comp}}

For a fair comparison, we pick two points with very similar mass spectra,
specifically ($\mzero=384~\GeV, \mth=44~\TeV$) in the mAMSB case and
($\mzero=455~\GeV, \mhalf=420~\GeV$) from the CMSSM\@.
These points have degenerate squark and gluino masses of about 980~\GeV,
lying near the border of the 95$\%$ ATLAS CMSSM exclusion limit in the equal
squark-gluino mass limit.
We display the two points' spectra and most likely decays in
Fig.~\ref{fig:spec}.   In the figure, the quasi-degenerate
lightest neutralino and lightest chargino are evident for the mAMSB point. 
\begin{figure}[htbp!]
\begin{center}
\subfigure[mAMSB point]{\includegraphics[height=0.35\textheight]{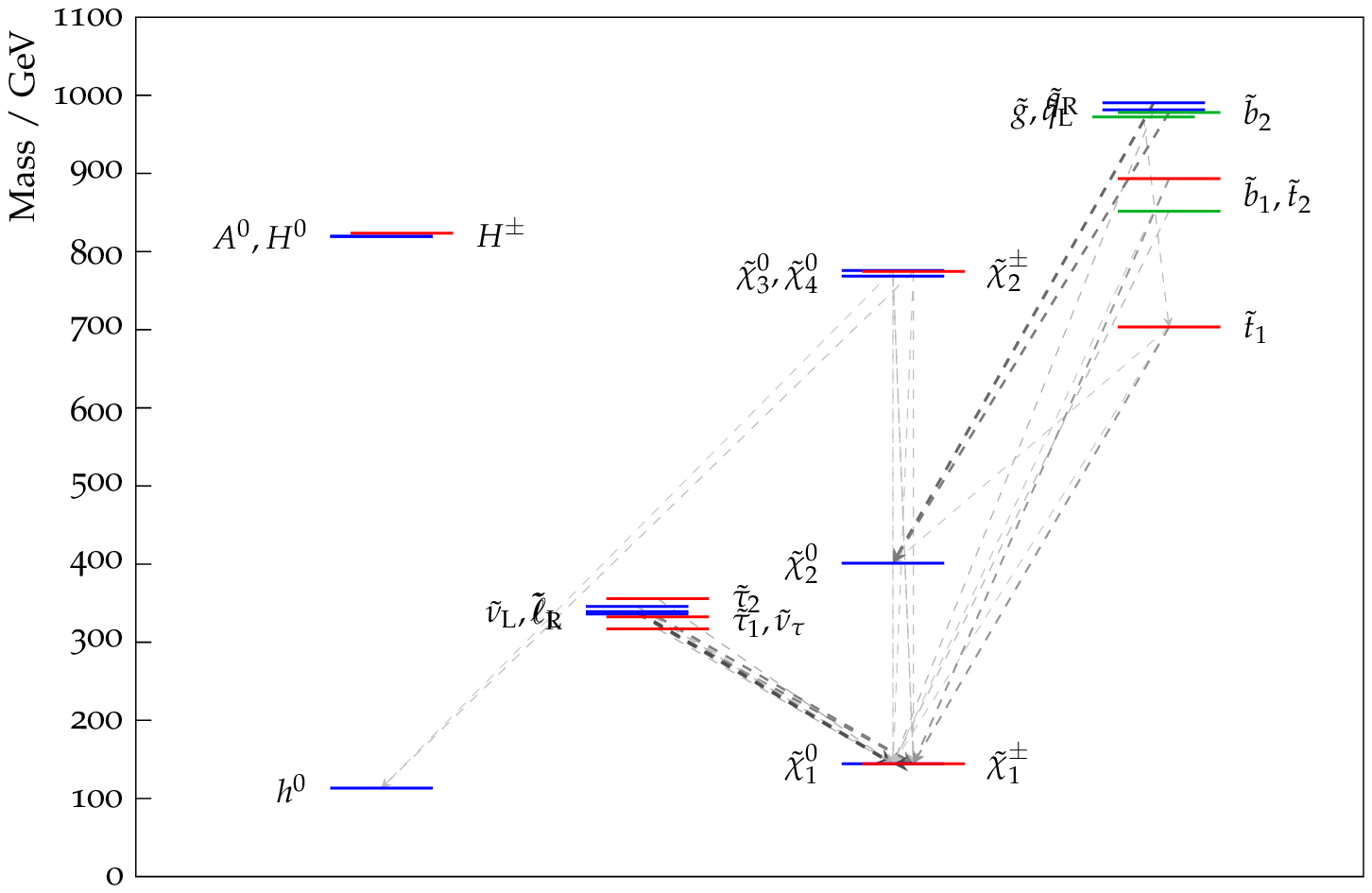}}
\subfigure[CMSSM point]{\includegraphics[height=0.35\textheight]{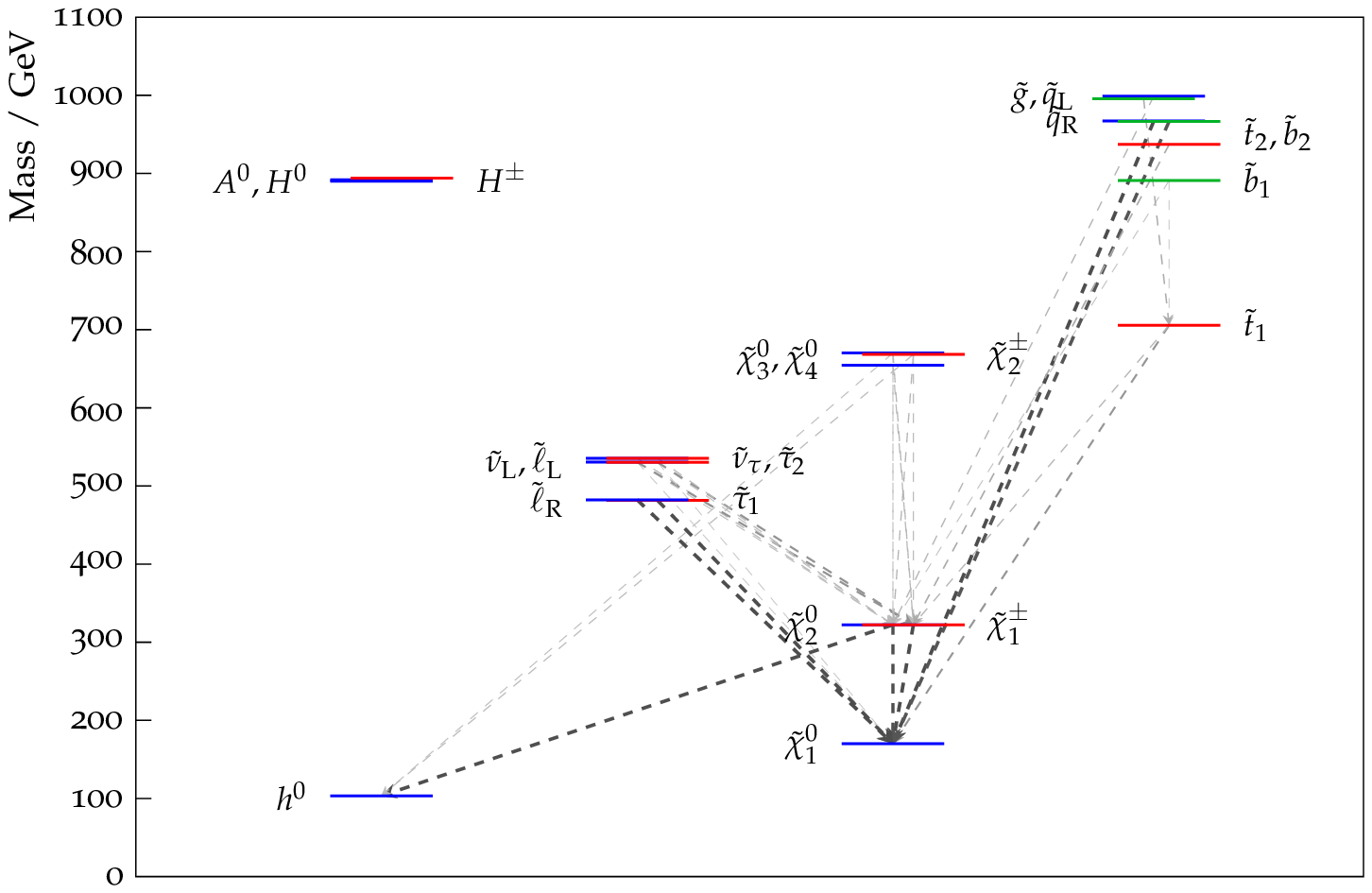}}
\end{center}
\caption{\label{fig:spec}Spectra and decays of the mAMSB and CMSSM model
  points studied. Only decays whose branching ratios are higher than 20$\%$
  are shown by the arrows. Both points have $\tan \beta=10$, $\mu>0$.  For the mAMSB point, we have $m_0=384$ GeV and $m_{3/2}=44$ TeV, whereas the
 CMSSM point has $m_0=455$,  $m_{1/2}=420$ and $A_0=0$. }
\end{figure}

\begin{figure}\begin{center}
\includegraphics[height=0.18\textheight, clip, trim = 0 0 0 1.25cm]{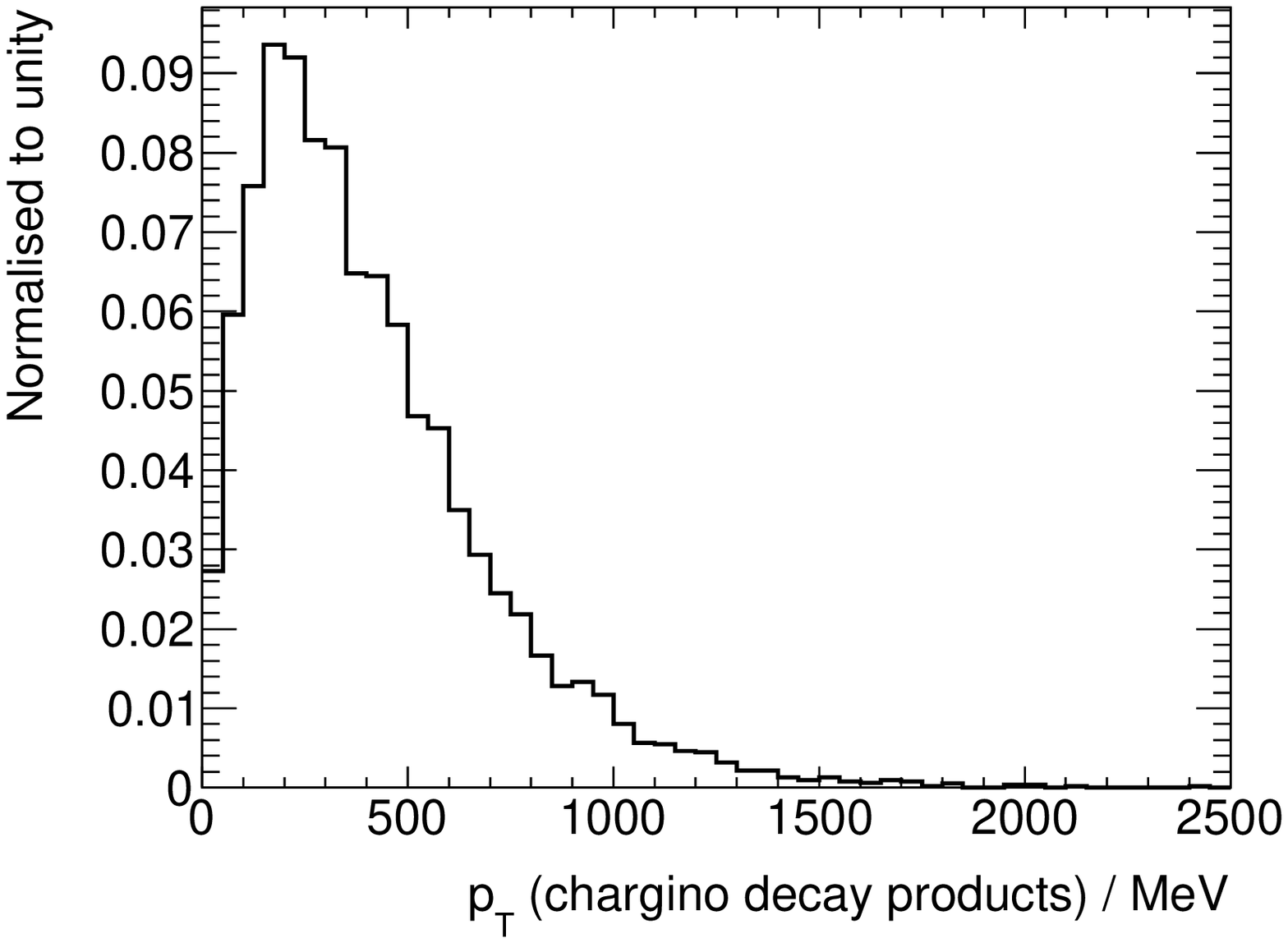}
\caption{$p_T$ spectrum of visible lightest chargino decay products in mAMSB,
  for $m_0=384$ GeV, $m_{3/2}=44$ TeV, $\tan \beta=10$ and $\mu>0$. 
  No detector simulation or kinematic selection is applied.
\label{fig:DK}}
\end{center}\end{figure}

We now turn to the characteristic lightest chargino decays in
mAMSB\@. In our simulated sample mAMSB point, 6002 lightest charginos were
produced, out of 10000 SUSY events. Of these, 24 decayed into a muon, 
138 into an electron and the rest into charged pions. 
We show the $p_T$ distribution of all lightest chargino 
decay products in Fig.~\ref{fig:DK}. The figure shows that all visible decay
products have  $p_T<2$  GeV; the LSP typically 
carries virtually all the momentum from the lightest chargino.
The SM decay products are difficult to distinguish from other soft particles
produced in LHC events, and would certainly require a dedicated analysis 
such as the one in Ref.~\cite{Barr:2002ex} in order to verify that they come
from the lightest chargino decay. For an analysis such as ours, these decays are
effectively invisible. 
However, hard leptons are available from other decay chains. For instance,
because the 
sleptons are between the $\chi_2^0$ and $\chi_1^0$ masses in the mAMSB point,
decays through the $\chi_2^0$ will more often lead to hard leptons, violating
the lepton veto and leading to less efficiency than in the CMSSM point for the
zero leptons channel. 

To illustrate the similarities and differences in the mAMSB and CMSSM models,
we compare a selection of relevant kinematic variables in
Fig.~\ref{fig:propsl}.
We wish to see if the ATLAS selections are efficient for mAMSB, or whether
the distributions suggest radically different cuts. 
No {\em a priori} kinematic selection is applied to these events, apart from
the basic object 
selections needed to conform with ATLAS variable definitions, such as requiring
each jet used in the \meff{} computation to have 40~\GeV{} in \pt{}.

It is seen that the kinematics of the two model points are reasonably
similar. For example, the $\ptmiss$ distributions of the two
models look remarkably similar, making the $\ptmiss$ cut approximately equally
efficient in each case. However, 
compared to the CMSSM point, the mAMSB point has less jets, the hardest jet is
softer on average and \meff{}~is smaller. These all make the corresponding
cuts somewhat 
{\em less efficient} in the case of mAMSB as compared to the CMSSM. 
The cuts on $\ptmiss/\meff$ are {\em more} efficient for the mAMSB point, but
in fact this effect is swamped by the less efficient cuts on
jets and from the lepton veto.
We also see that $M_{T2}$ tends to be smaller for the
mAMSB point because of the softer jets \citemttwo{}. 
Although this variable is not used in the present
search, it has similar search power to other methods, but in some cases can
discover generic MSSM parameter points when the usual $\meff$, $\ptmiss$
searches cannot~\cite{Allanach:2011ej}. 
We therefore advocate its inclusion as part of the searches, even though
a lower cut of a
few hundred GeV looks to be slightly
more efficient for the CMSSM model point than the mAMSB model point examined.

We may understand these kinematic differences as follows:
the sparticle cascade decay chains starting from gluinos or squarks, 
feature decays through lightest charginos prominently in both the CMSSM and
mAMSB points.  
In mAMSB, the lightest charginos are invisible to our analysis as explained
above, whereas at the CMSSM point, they decay to $W \chi_1^0$, so the $W$
leads to additional jets (any that decay leptonically are likely to be vetoed),
contributing to the jet multiplicity and \meff. A softer \meff{} for mAMSB then
leads to a more highly peaked \meff/\ptmiss{} ratio. There is also a larger
cross-section for weak gaugino/strongly interacting SUSY particle production
in mAMSB, making \meff{} softer. More of the squarks decay through $\chi_2^0$
in mAMSB compared to mAMSB, and since it is heavier, this reduces the $p_T$ of
the (typically hardest) jet involved in the decay. 

\begin{figure}\begin{center}
\subfigure{\includegraphics[width=0.4\textwidth, clip, trim = 0 0 2.75cm 0]{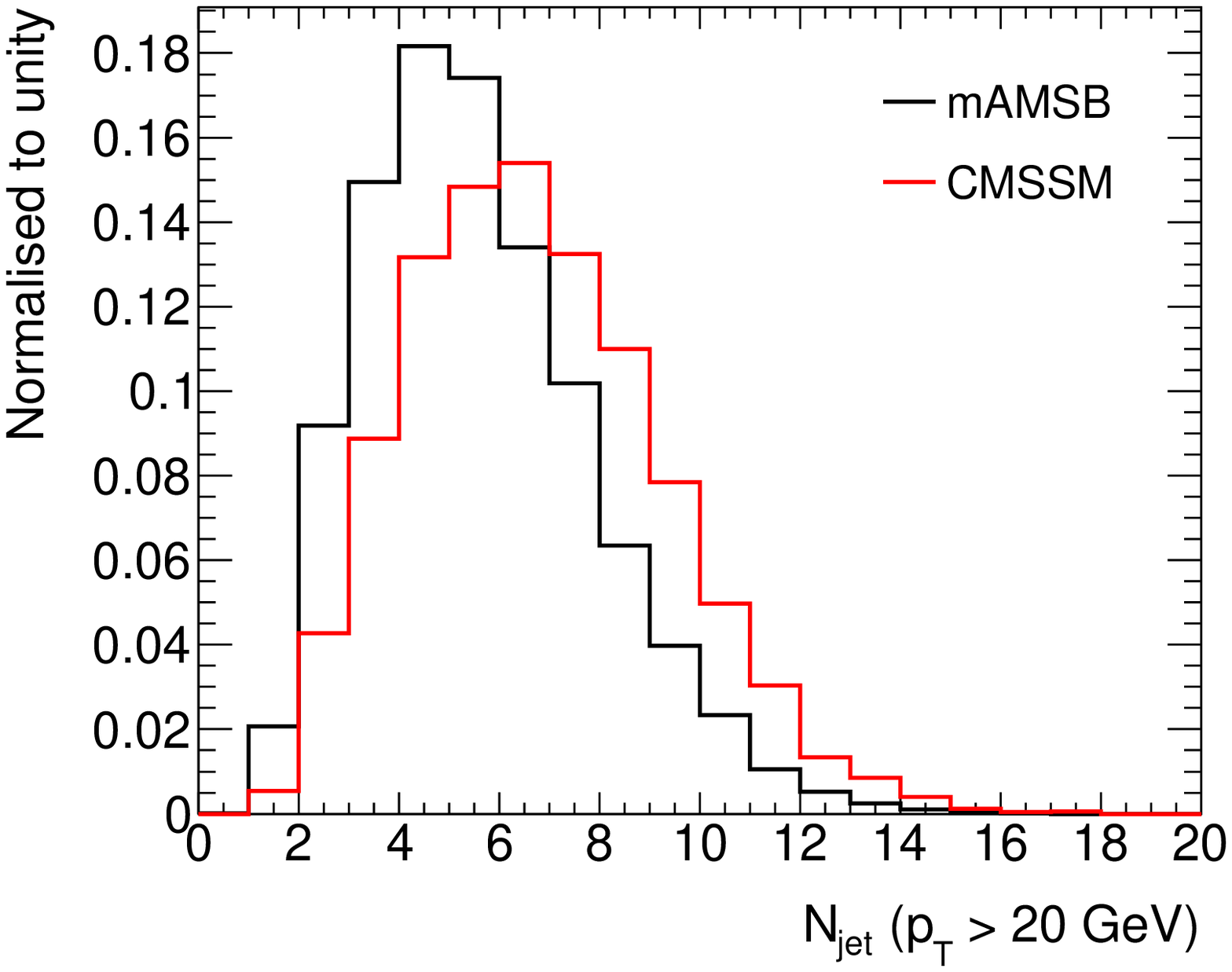}}
~\subfigure{\includegraphics[width=0.4\textwidth, clip, trim = 0 0 2.75cm 0]{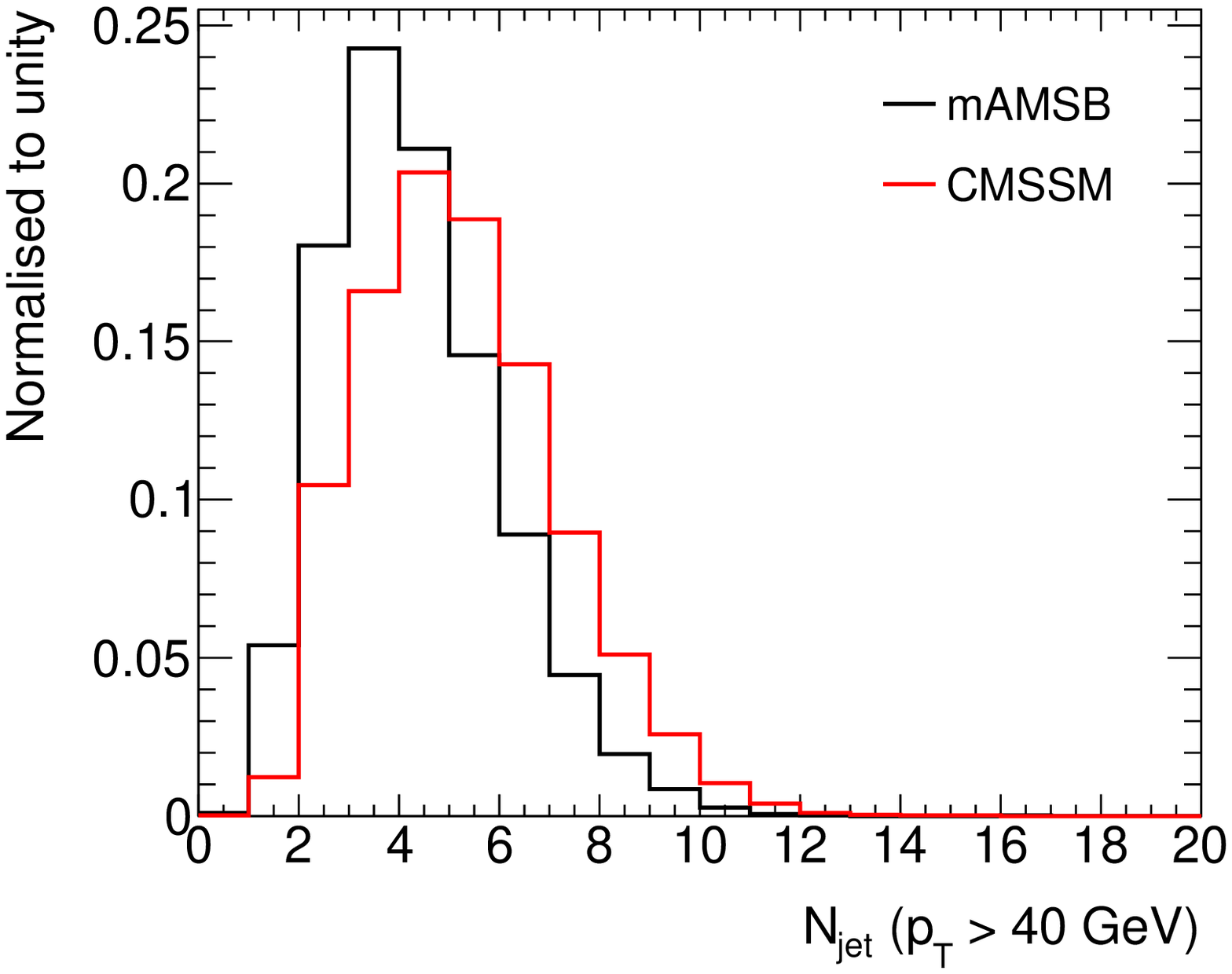}}

\subfigure{\includegraphics[width=0.32\textwidth, clip, trim = 0 0 2.75cm 0]{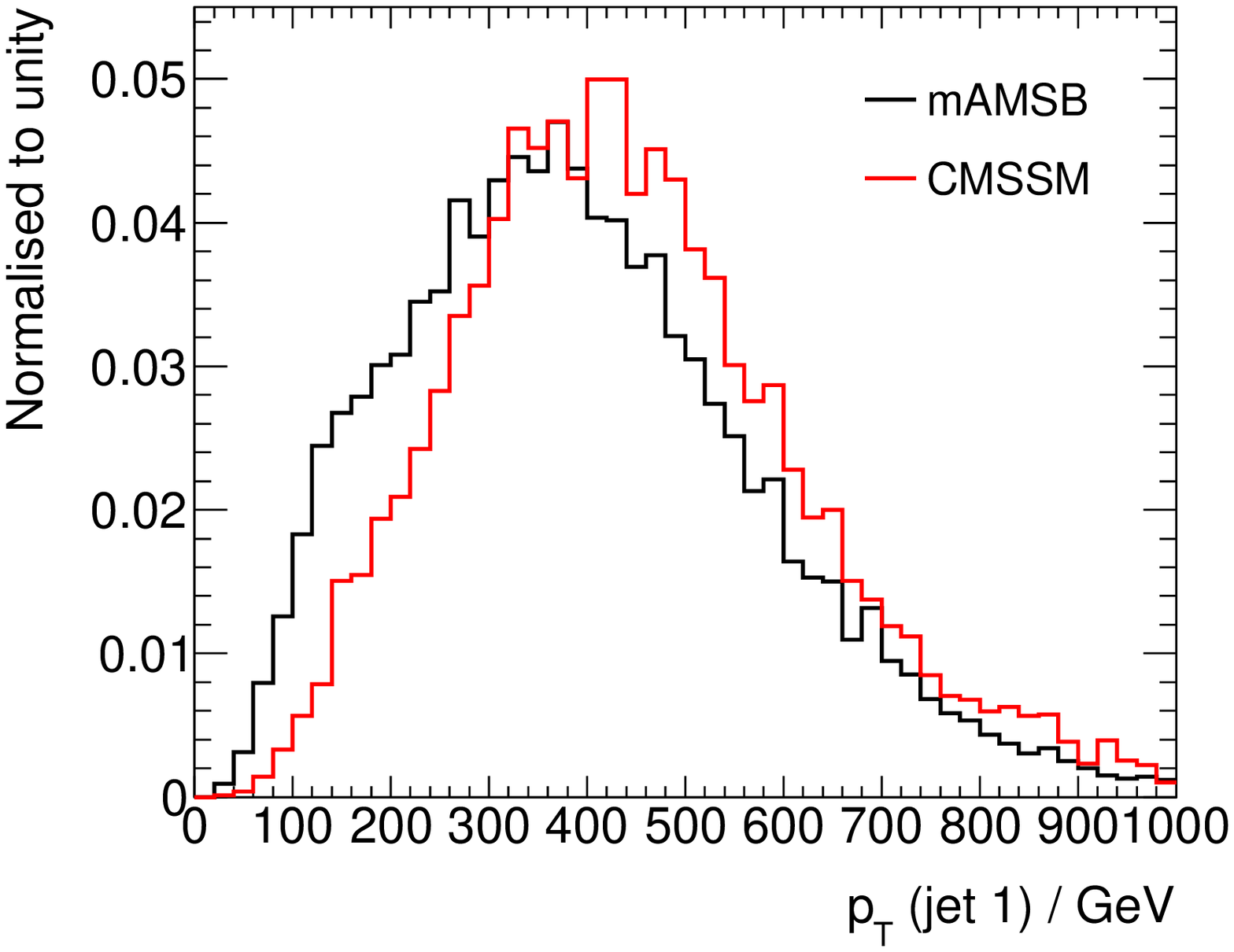}}
~\subfigure{\includegraphics[width=0.32\textwidth, clip, trim = 0 0 2.75cm 0]{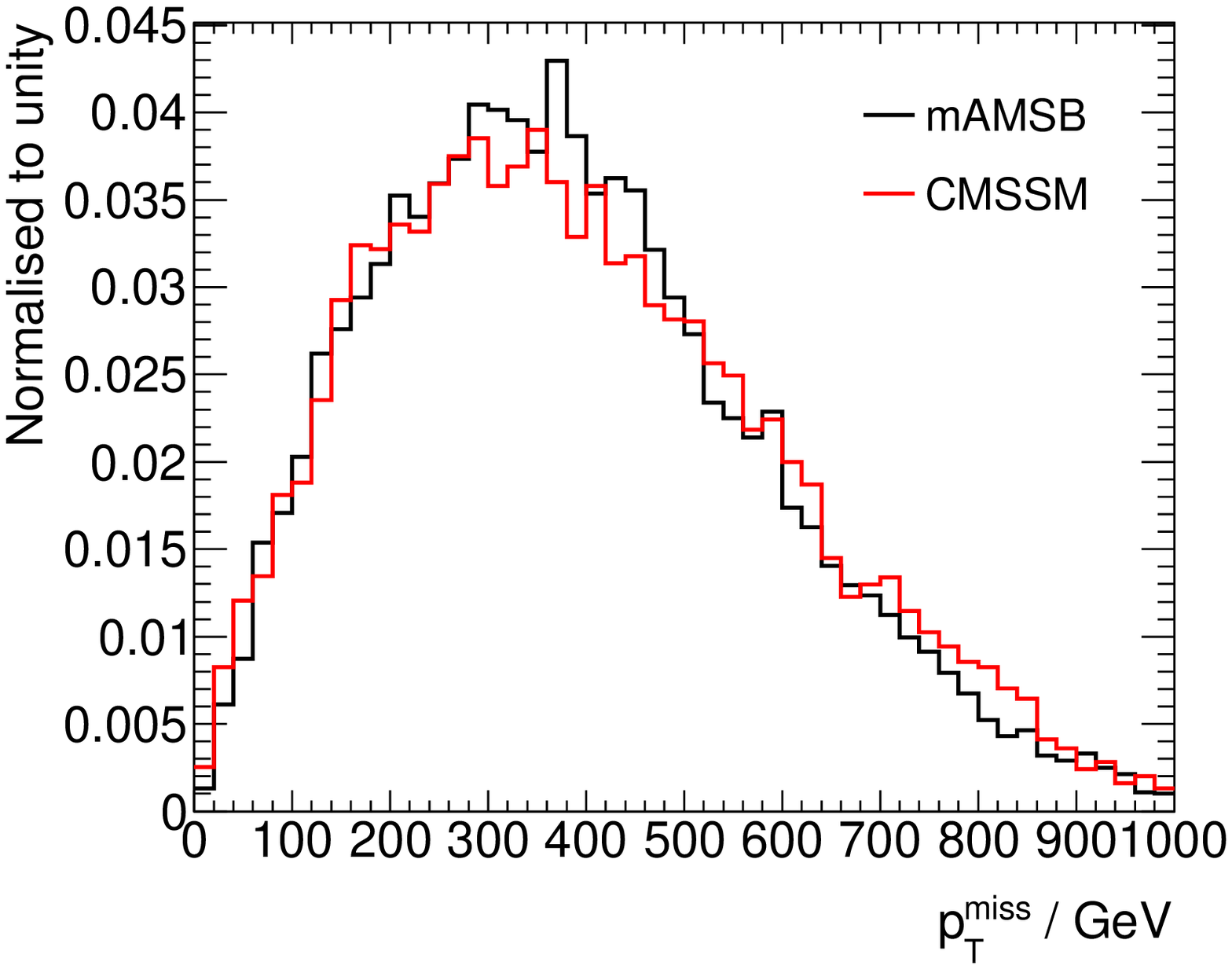}}
~\subfigure{\includegraphics[width=0.32\textwidth, clip, trim = 0 0 2.75cm 0]{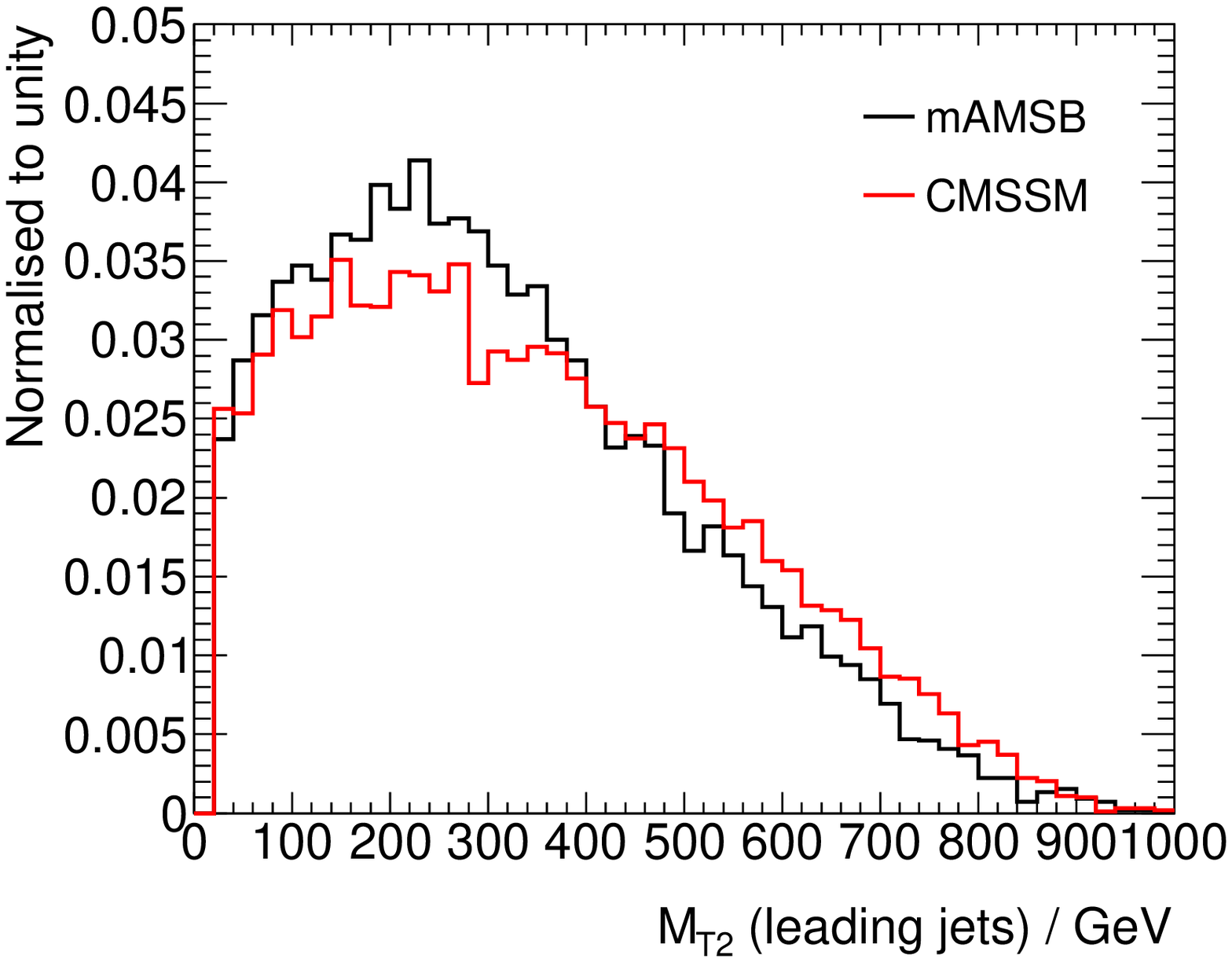}}

\subfigure{\includegraphics[width=0.32\textwidth, clip, trim = 0 0 2.75cm 0]{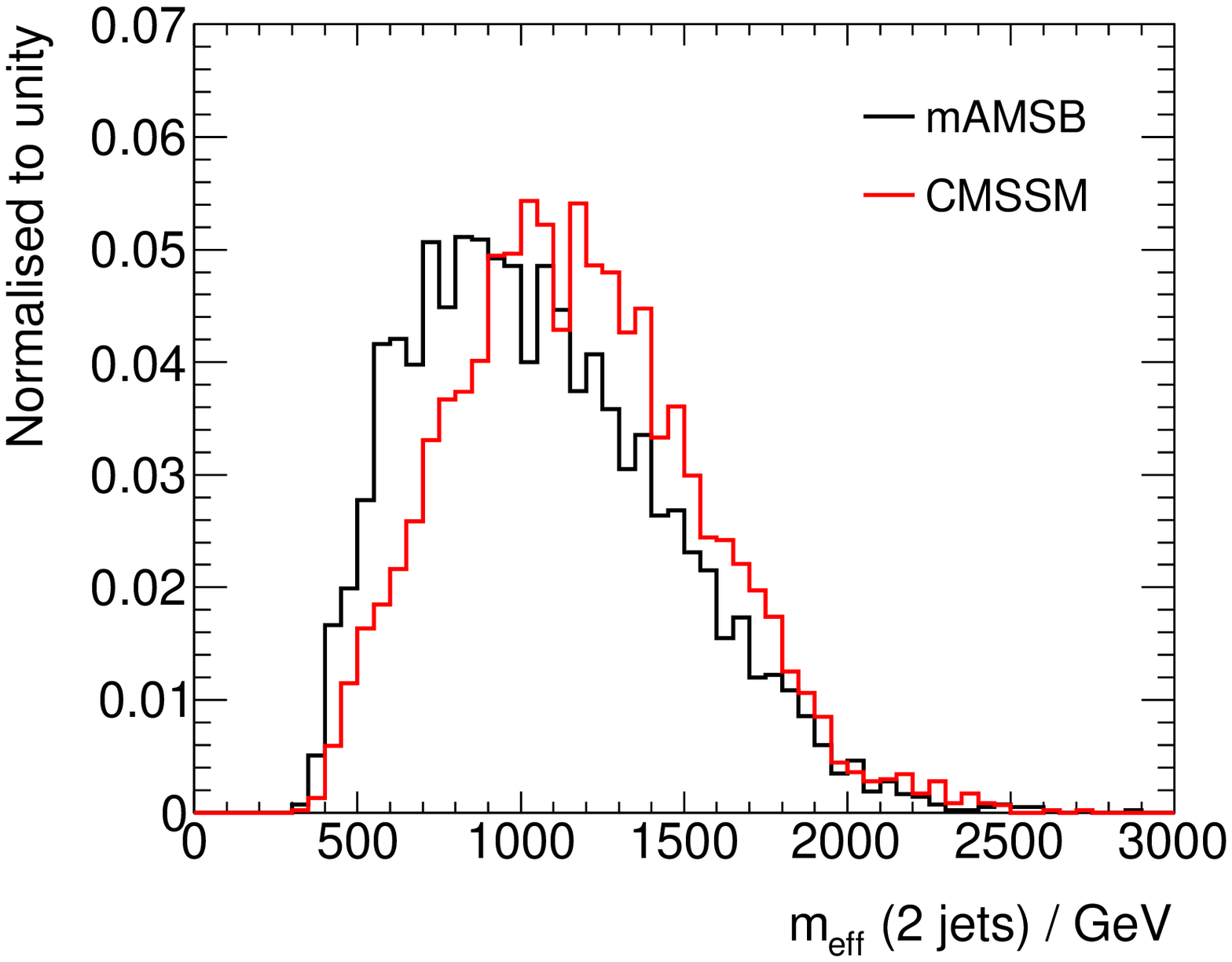}}
~\subfigure{\includegraphics[width=0.32\textwidth, clip, trim = 0 0 2.75cm 0]{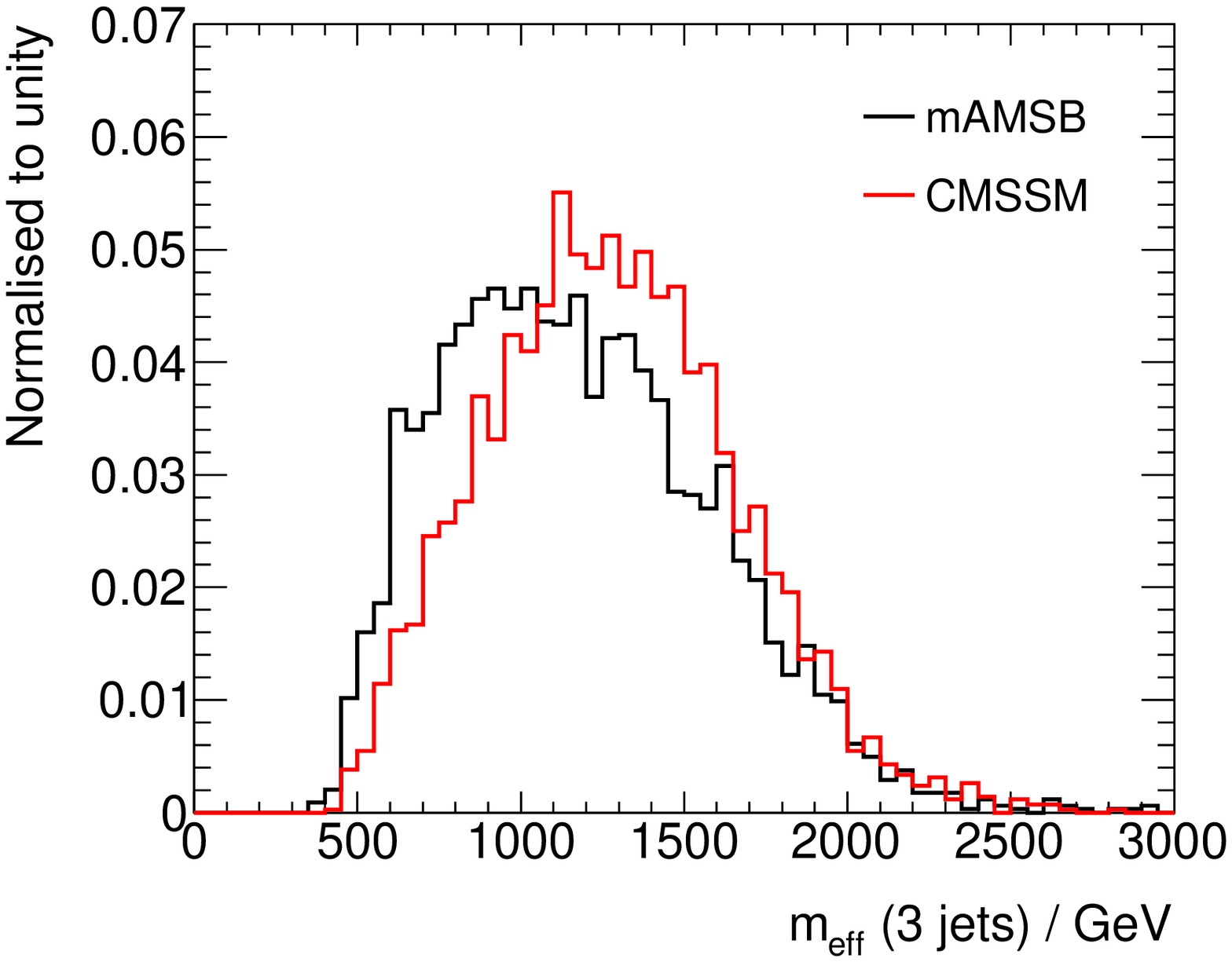}}
~\subfigure{\includegraphics[width=0.32\textwidth, clip, trim = 0 0 2.75cm 0]{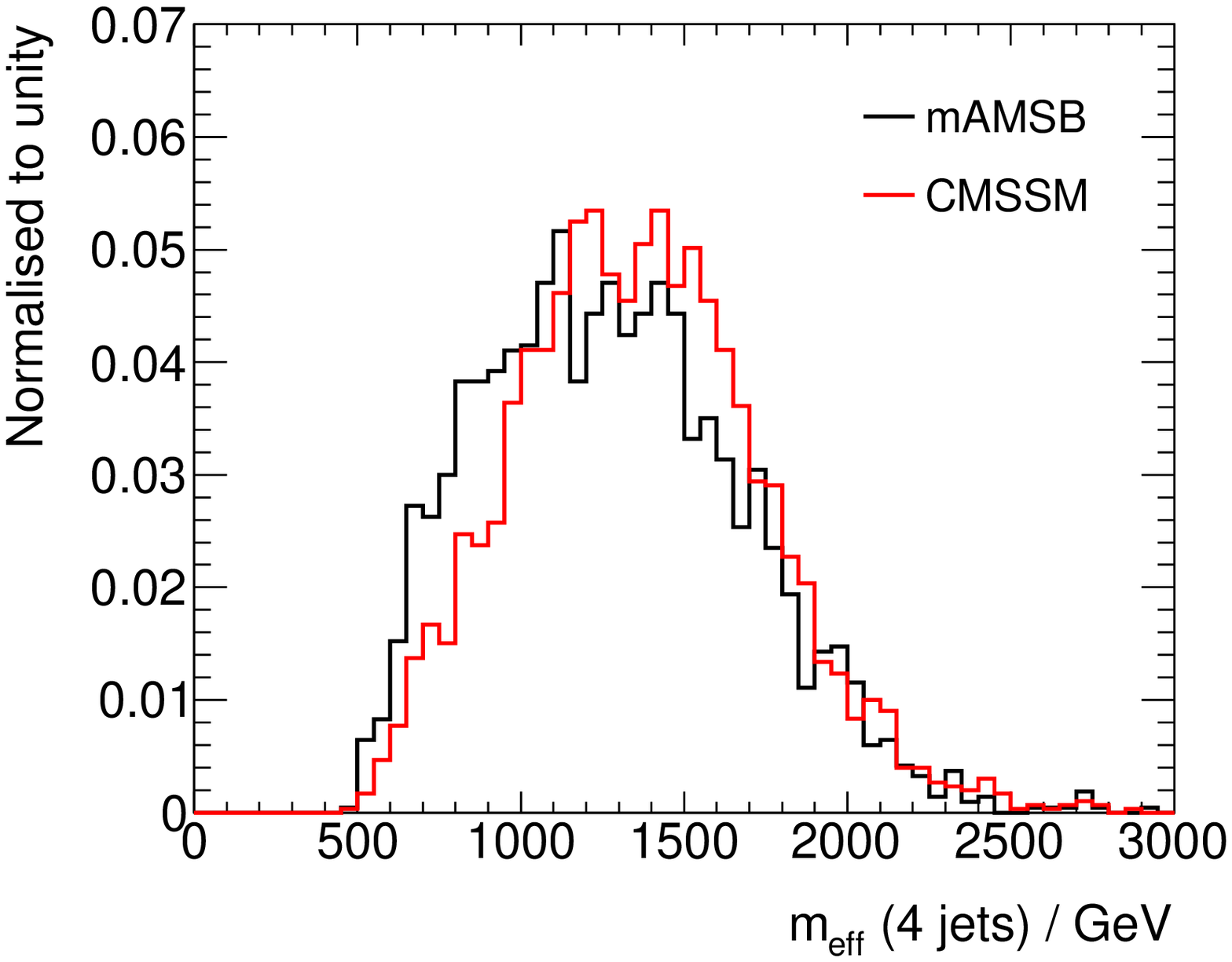}}

\subfigure{\includegraphics[width=0.32\textwidth, clip, trim = 0 0 2.75cm 0]{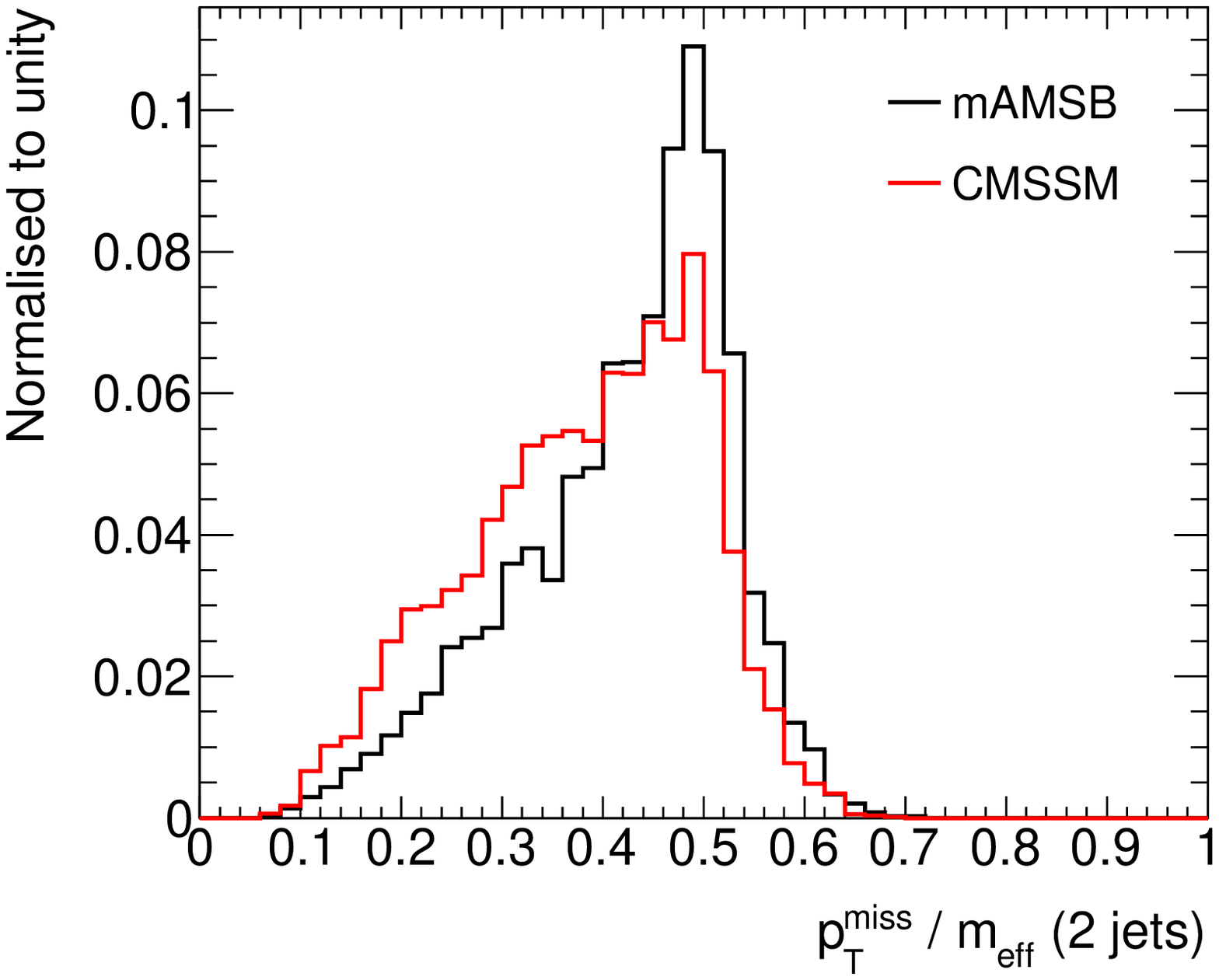}}
~\subfigure{\includegraphics[width=0.32\textwidth, clip, trim = 0 0 2.75cm 0]{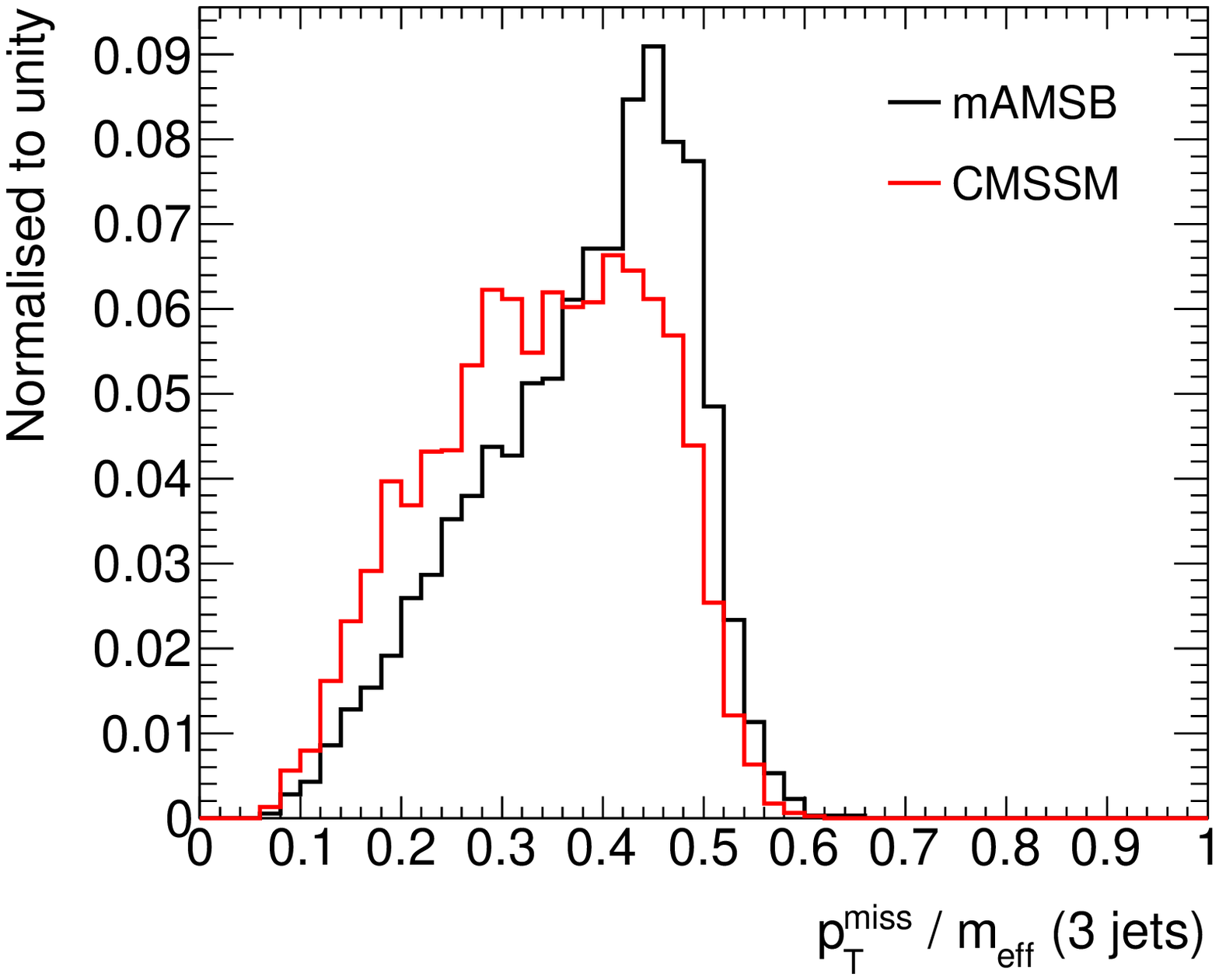}}
~\subfigure{\includegraphics[width=0.32\textwidth, clip, trim = 0 0 2.75cm 0]{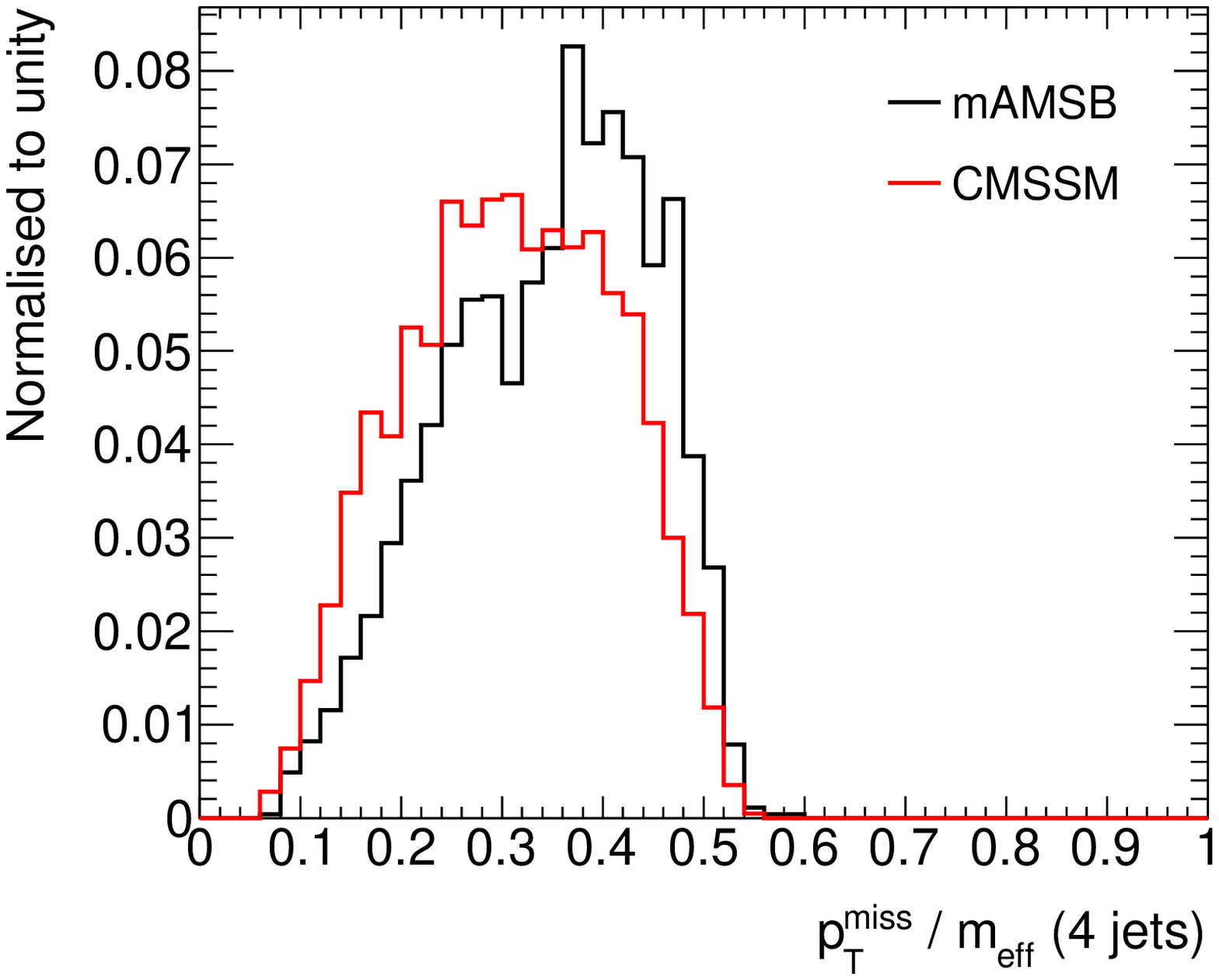}}

\caption{Important kinematic distributions of the signals for mAMSB and CMSSM
  sample model   points for $\tan \beta=10$ and $\mu>0$.
 For the mAMSB point, we have $m_0=384$ GeV and $m_{3/2}=44$ TeV, whereas the
 CMSSM point has $m_0=455$,  $m_{1/2}=420$ and $A_0=0$. 
 Only minimal kinematic cuts are applied, i.e. requiring two, three or four
 jets with 
 \pt{}$>$ 40~\GeV\ for the \meff{} and \ptmiss{}/\meff{} distributions,
 as is appropriate.
\label{fig:propsl}}
\end{center}\end{figure}

\subsection{mAMSB Scan: Properties of SUSY Events}

\begin{figure}[htbp]
\begin{center}

\subfigure{\includegraphics[width=0.32\textwidth]{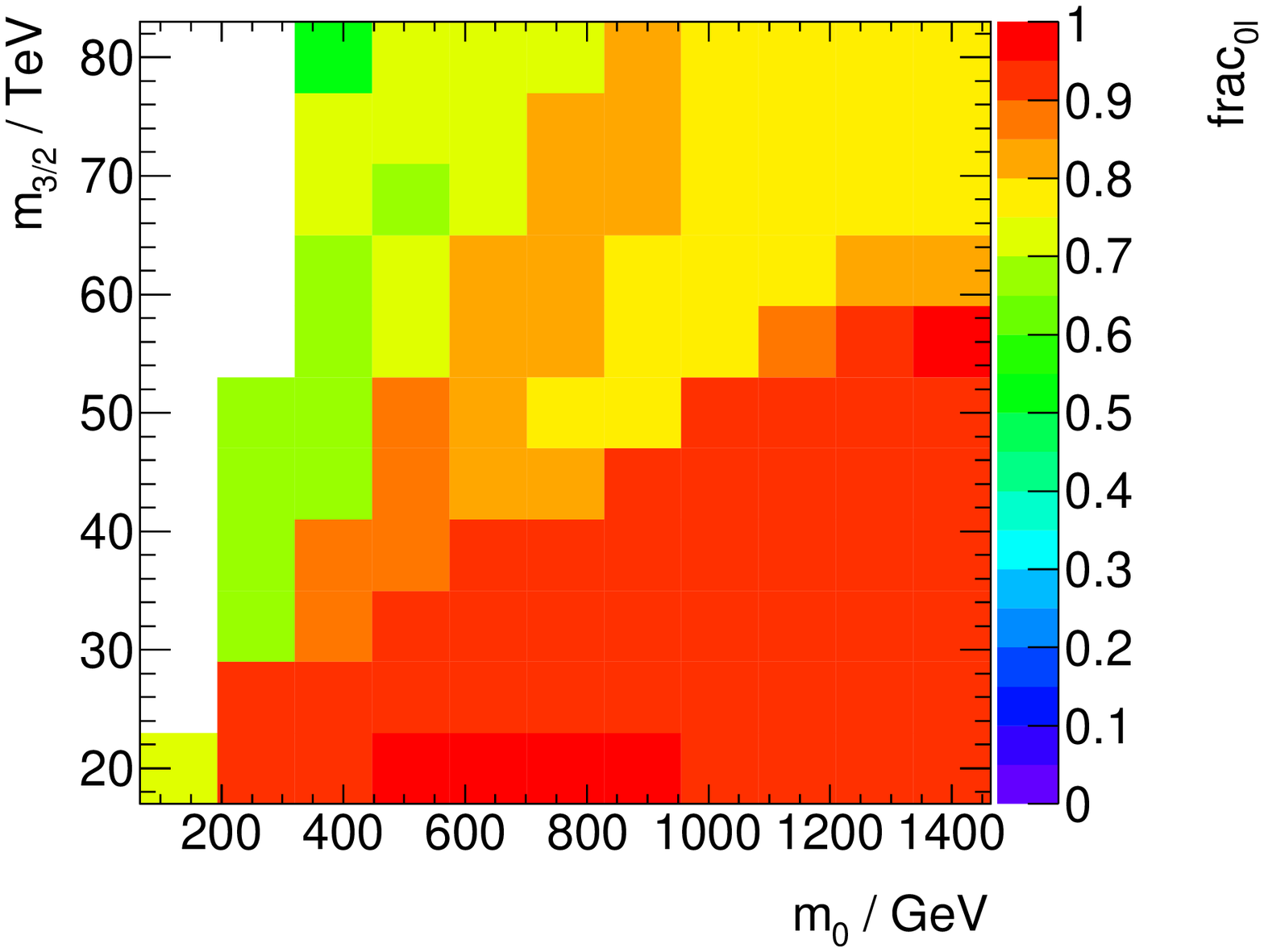}}
~\subfigure{\includegraphics[width=0.32\textwidth]{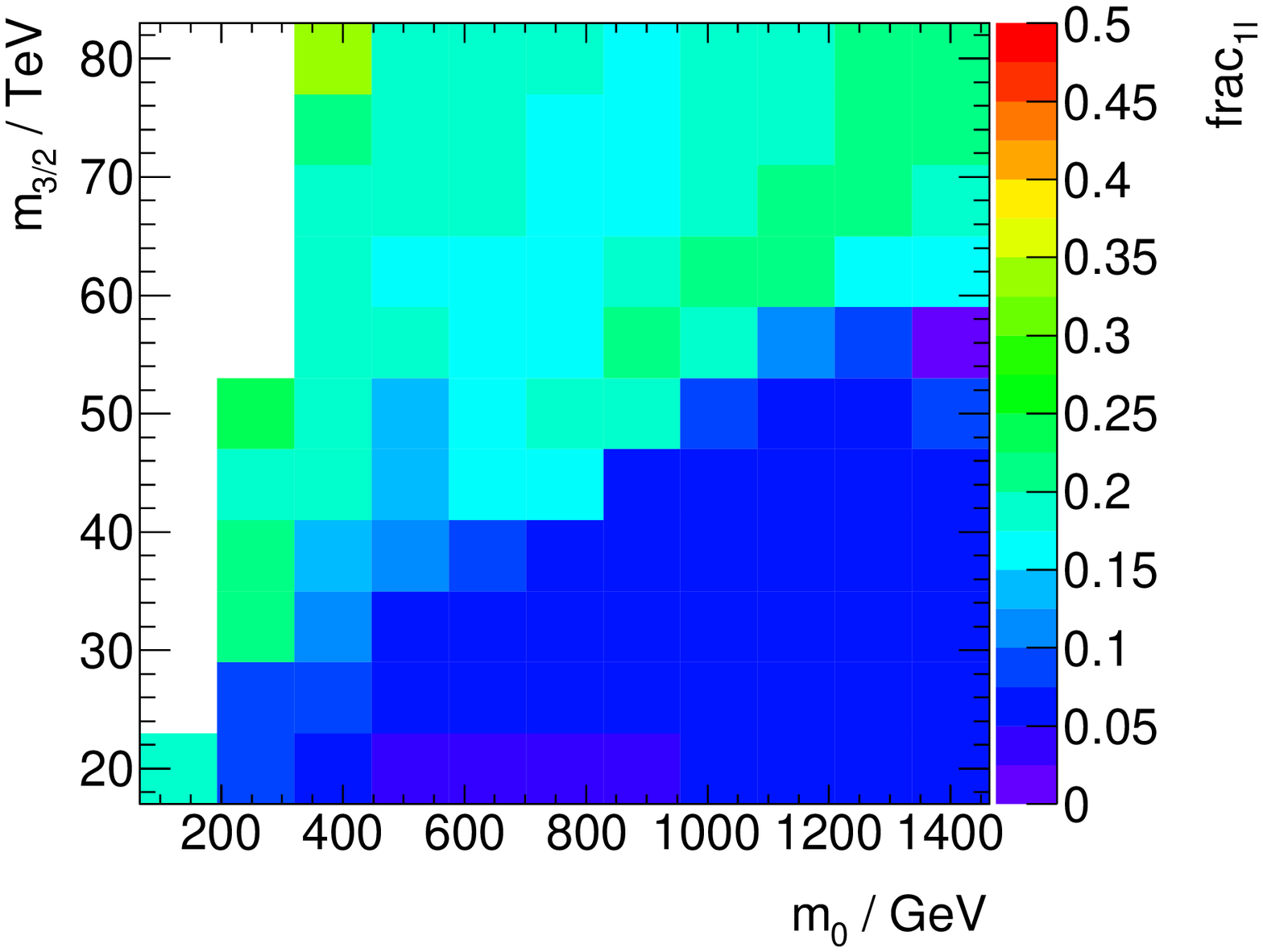}}
~\subfigure{\includegraphics[width=0.32\textwidth]{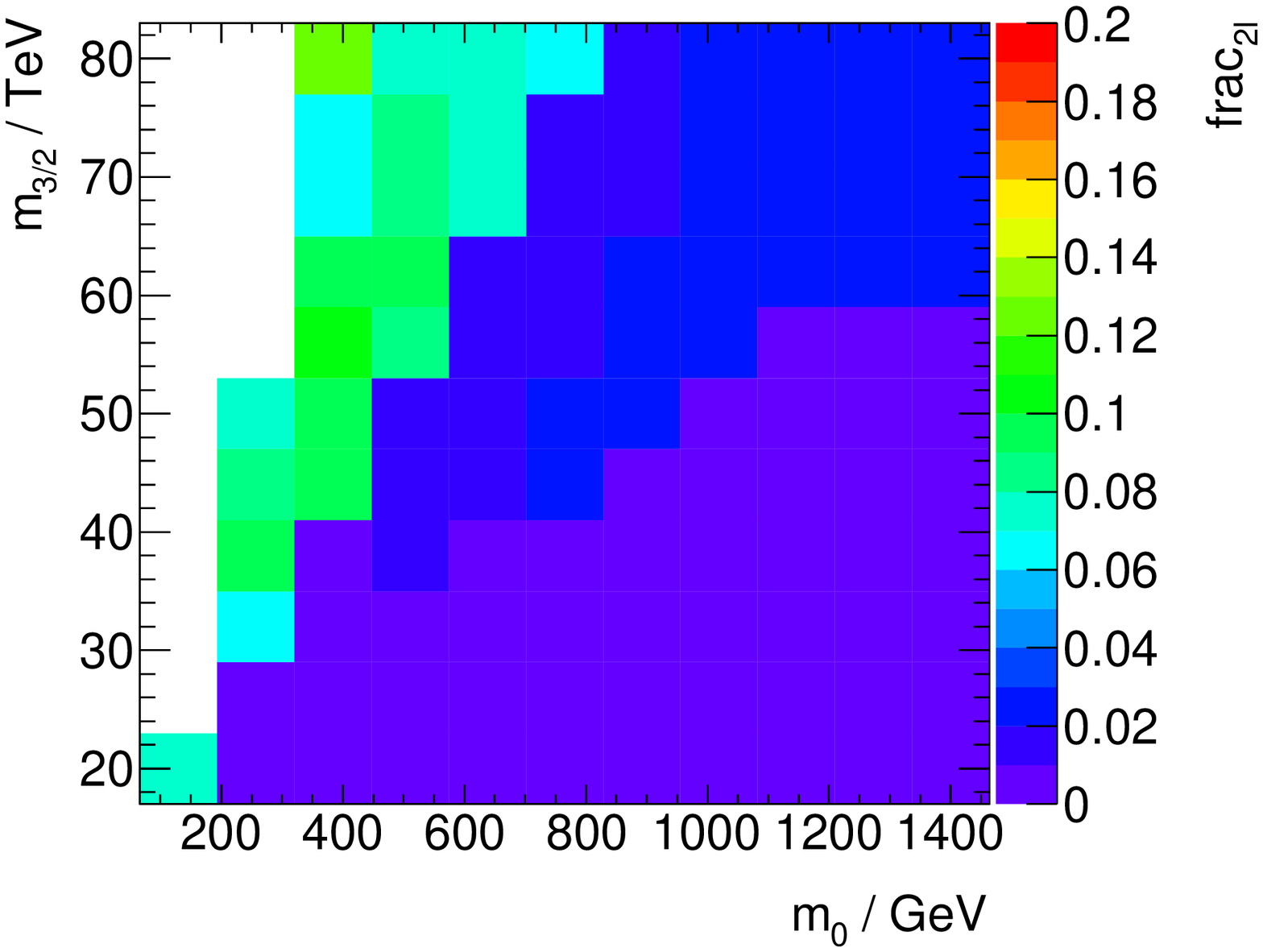}}

\caption{Fraction of events with 0, 1 or 2 hard isolated leptons in the mAMSB
  parameter space considered in this paper: $\tan \beta=10$, $\mu>0$. Leptons 
  with $\pt > 20~\GeV$ are considered to be hard and isolated if they are not inside a
  jet (also with $\pt > 20~\GeV$). No additional kinematic selection is applied. Note the
  different z-axis scales in the three plots.
\label{fig:AMSBleptons}}
\end{center}
\end{figure}

\begin{figure}[htbp]
\begin{center}

\subfigure[2 jets]{\includegraphics[width=0.49\textwidth]{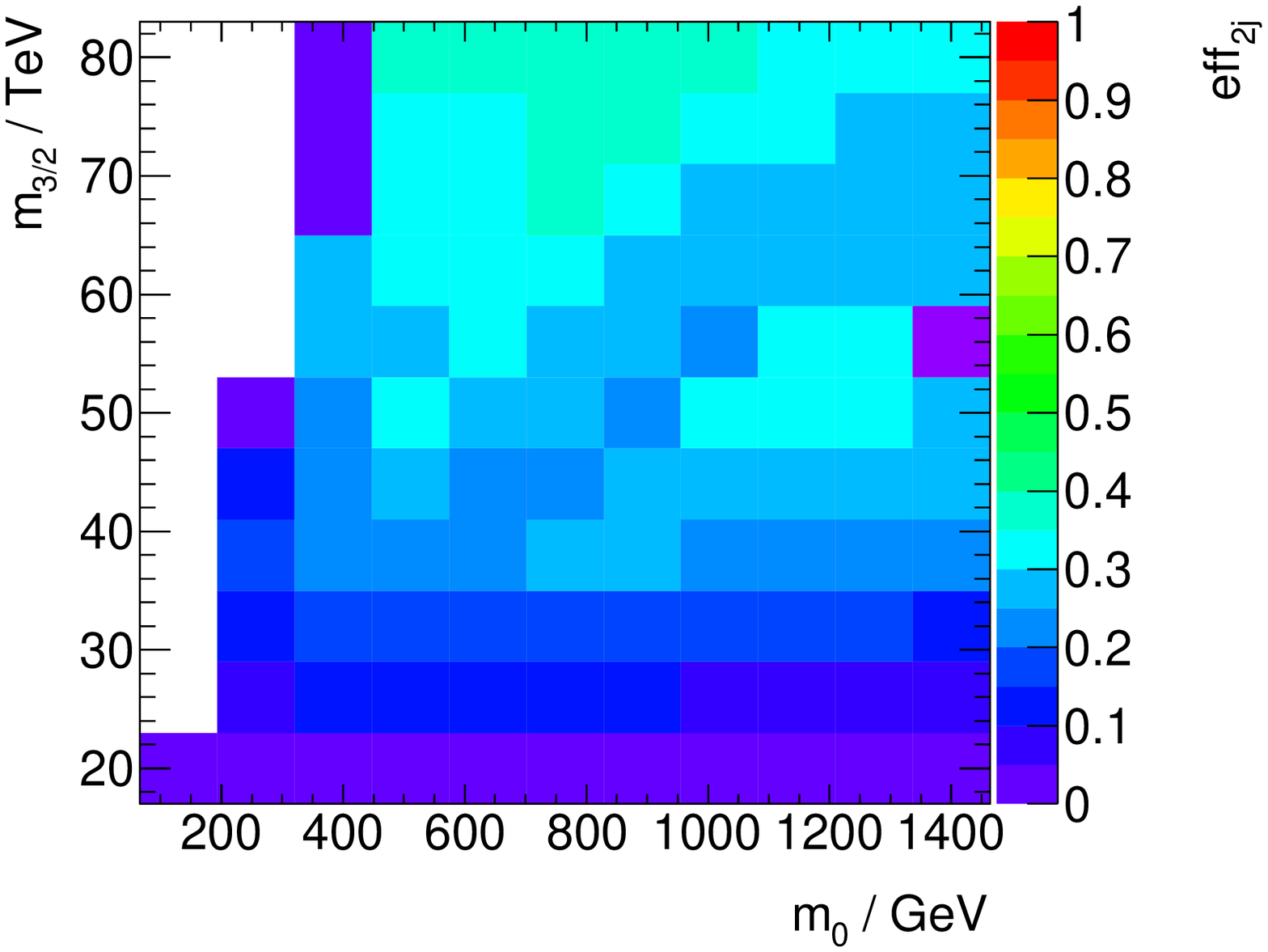}}
~\subfigure[3 jets]{\includegraphics[width=0.49\textwidth]{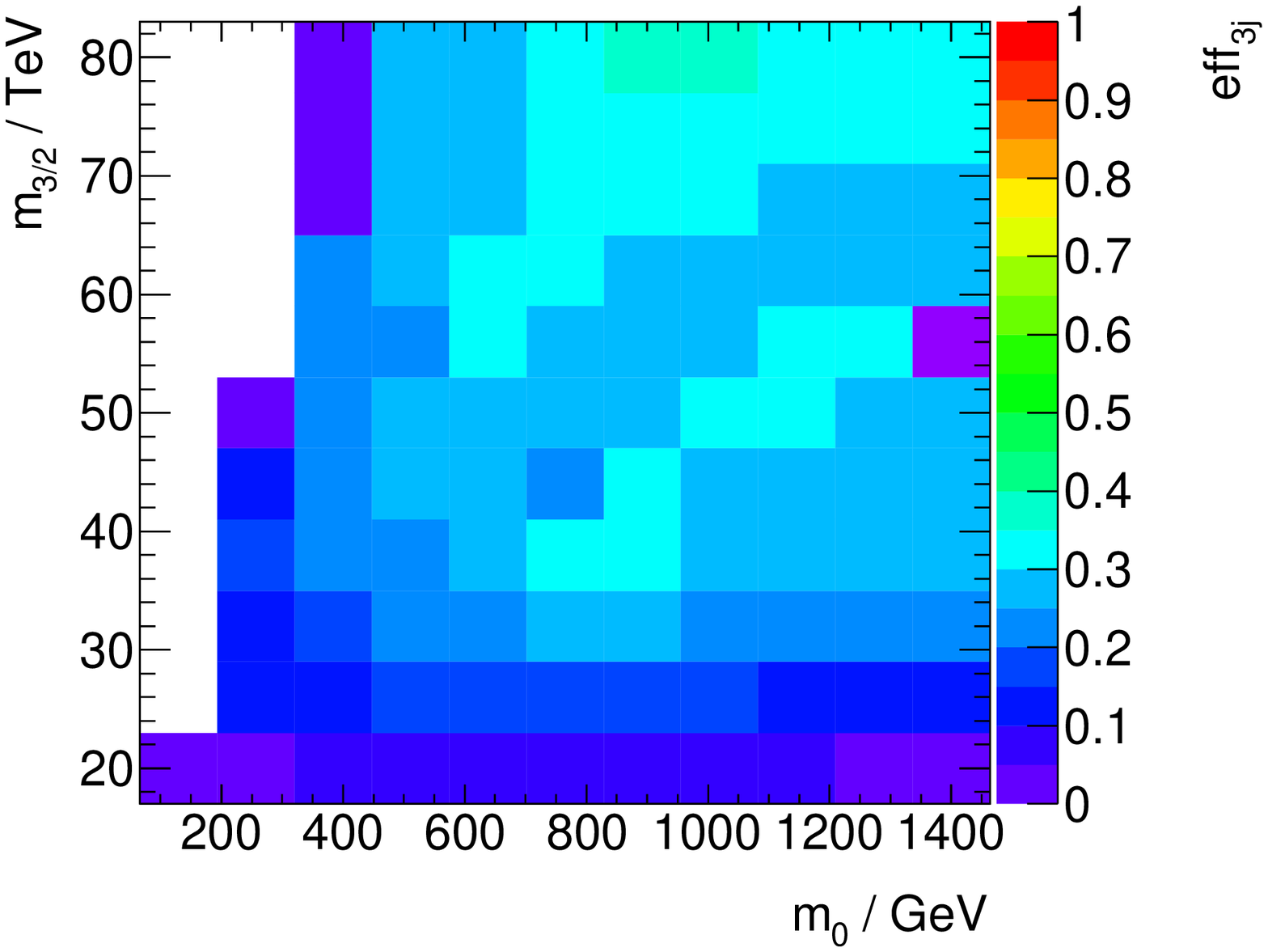}}

\subfigure[4 jets]{\includegraphics[width=0.49\textwidth]{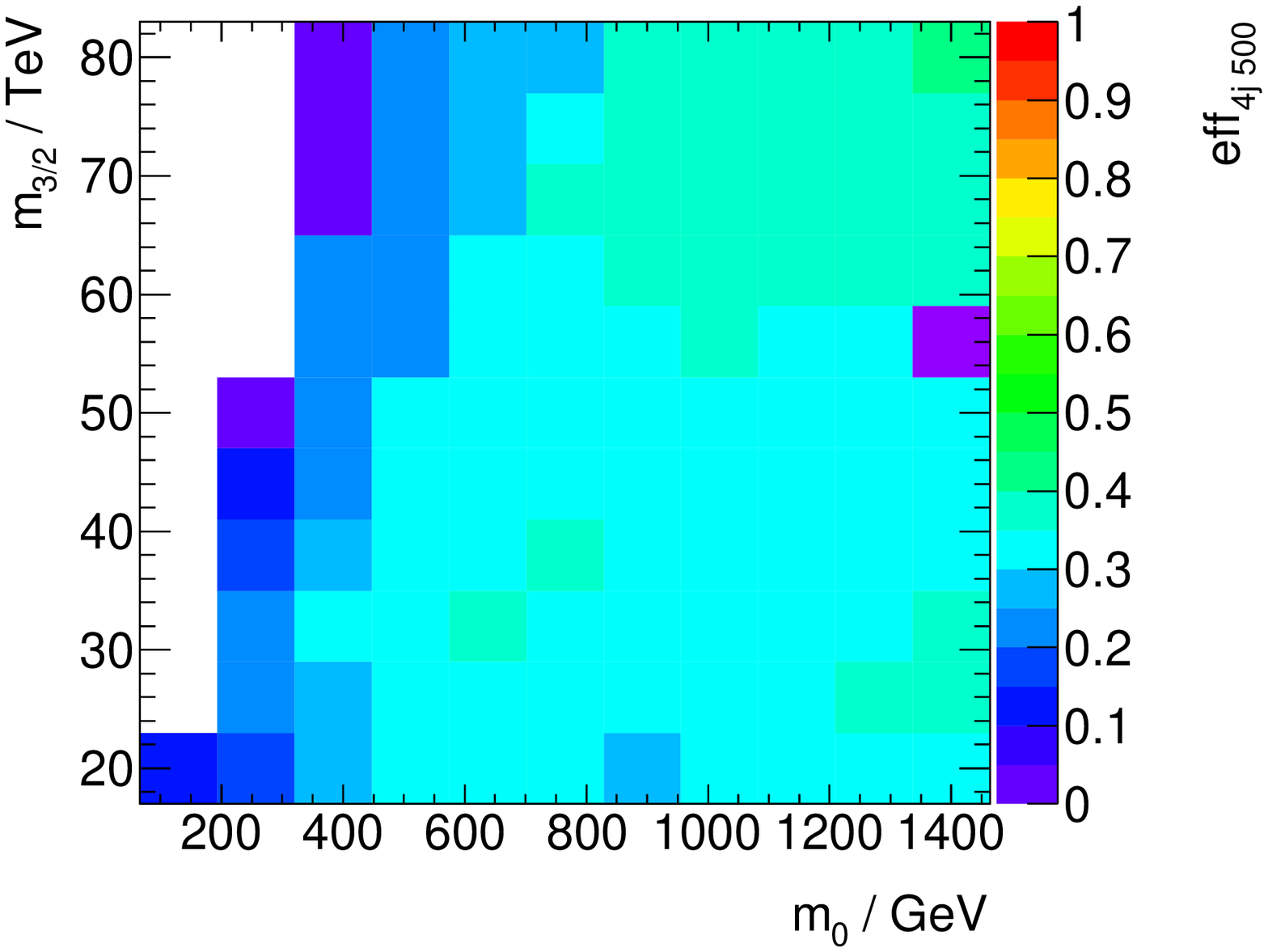}}
~\subfigure[4 jets']{\includegraphics[width=0.49\textwidth]{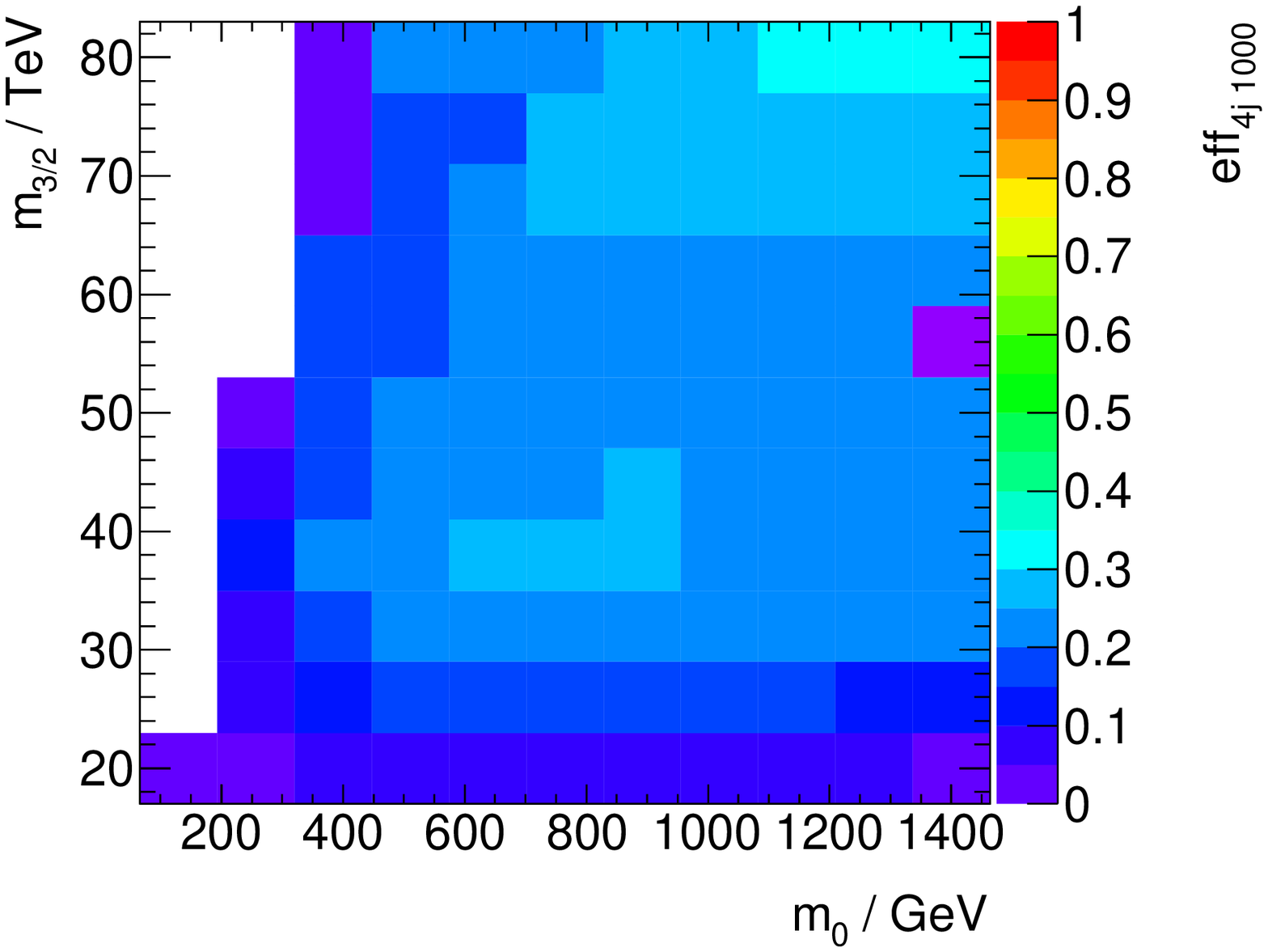}}

\subfigure[high mass]{\includegraphics[width=0.49\textwidth]{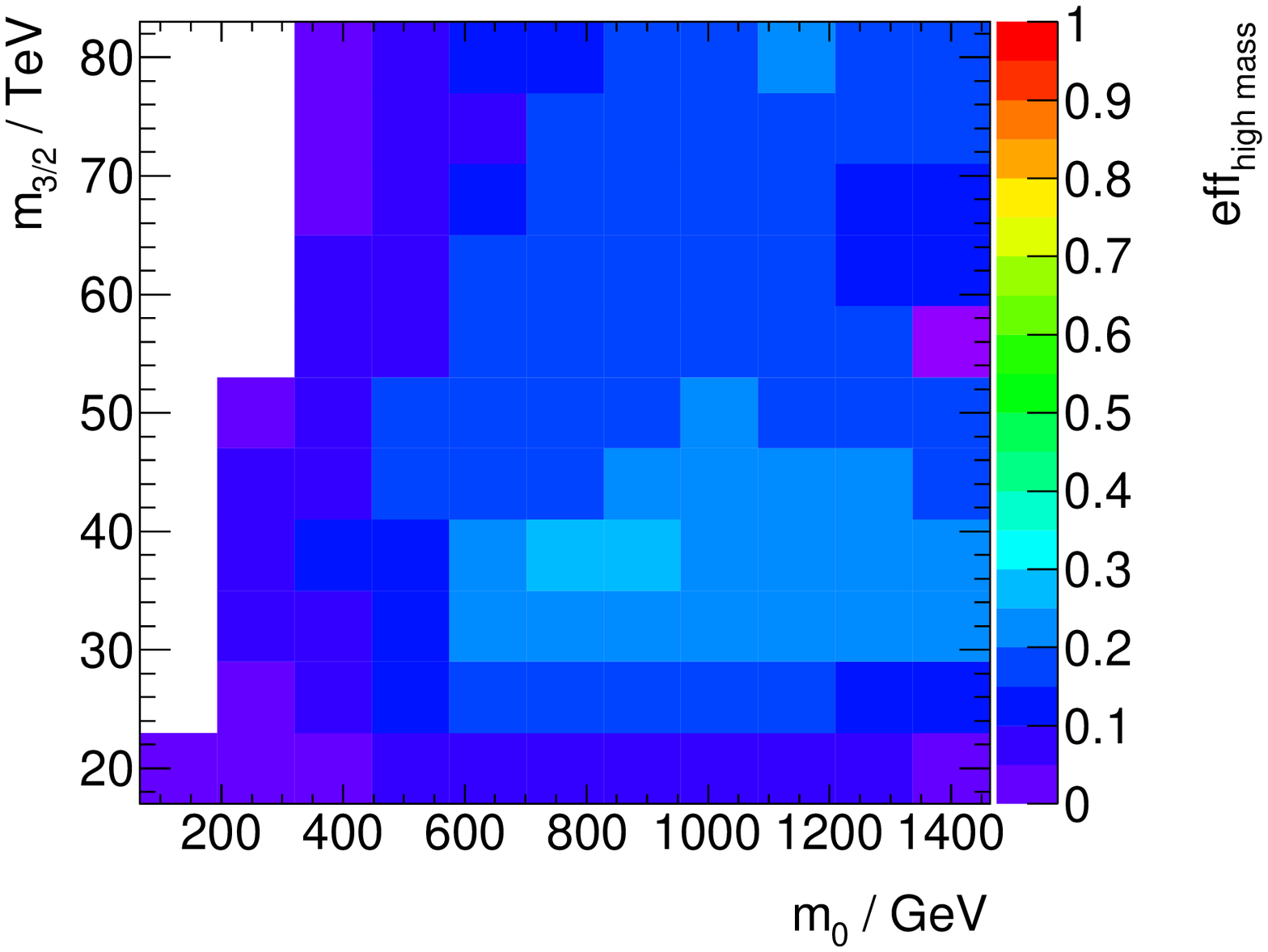}}

\caption{
  Efficiency of ATLAS signal selections in the mAMSB $\mzero-\mth$ plane.
  The flat correction factor for the missing calorimeter regions is not applied,
  as it adds no information about the physics of the signal models.
}
\label{fig:AMSBefficiency}
\end{center}
\end{figure}

\begin{figure}[htbp]
\begin{center}

\subfigure[2 jets]{\includegraphics[width=0.49\textwidth]{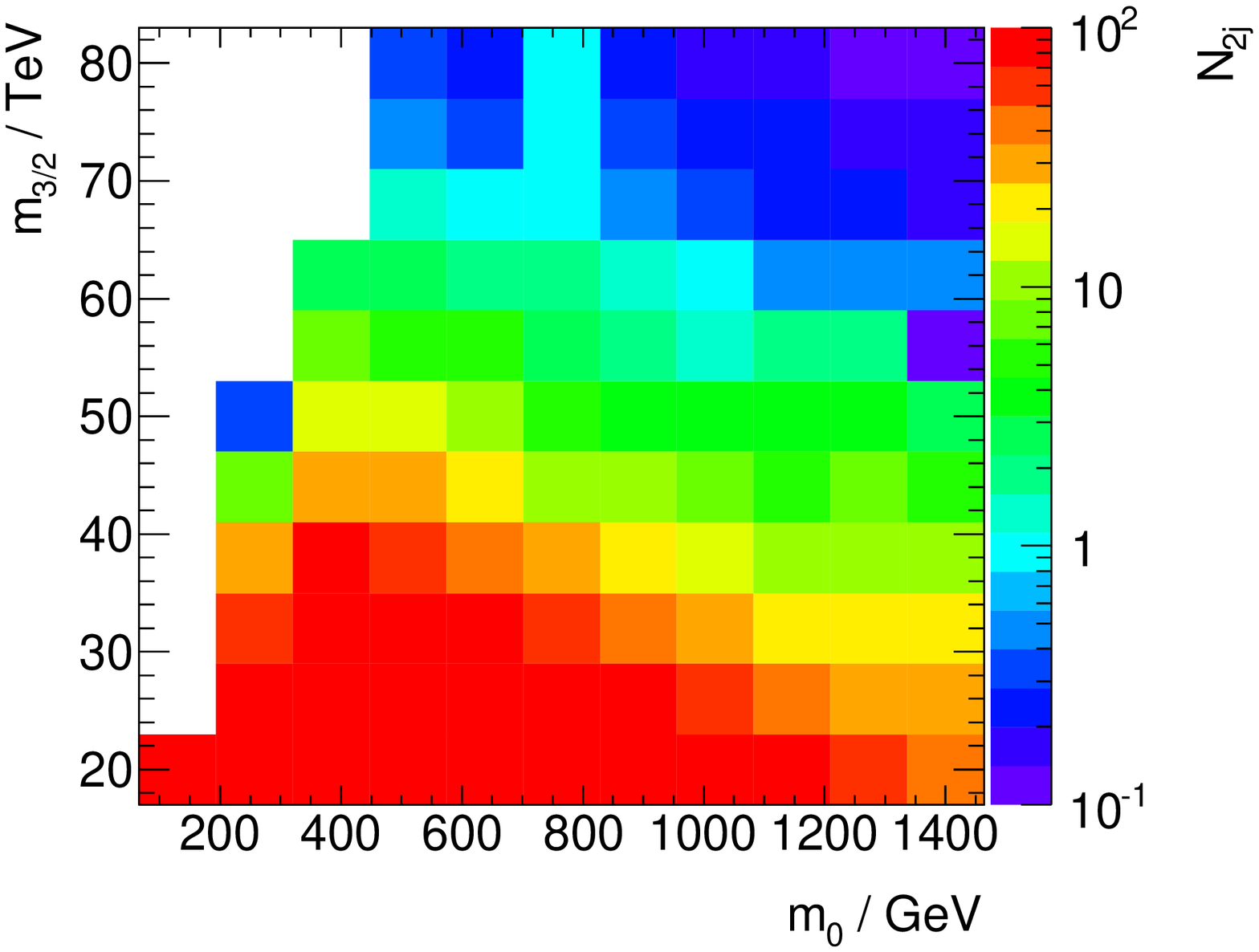}}
~\subfigure[3 jets]{\includegraphics[width=0.49\textwidth]{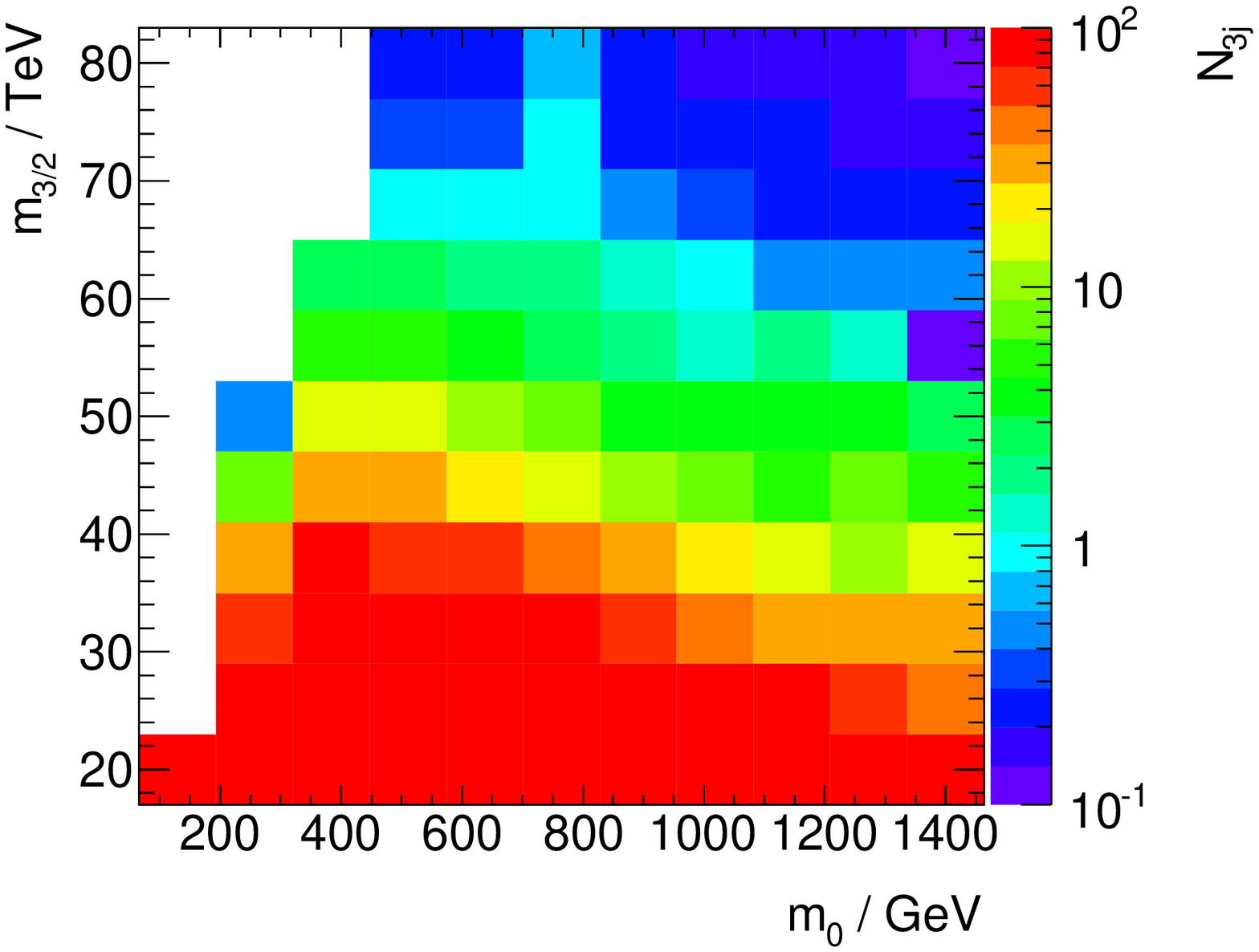}}

\subfigure[4 jets]{\includegraphics[width=0.49\textwidth]{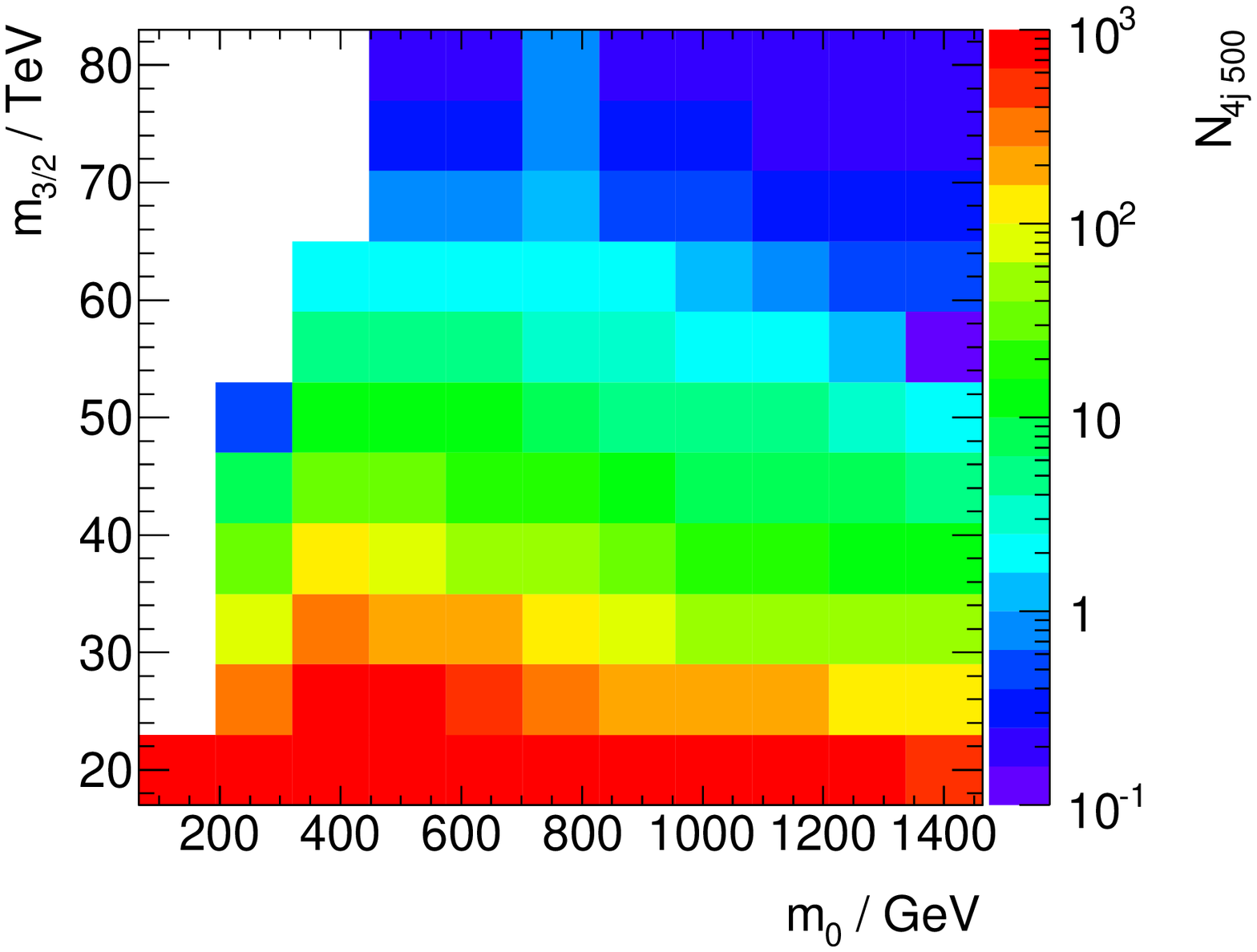}}
~\subfigure[4 jets']{\includegraphics[width=0.49\textwidth]{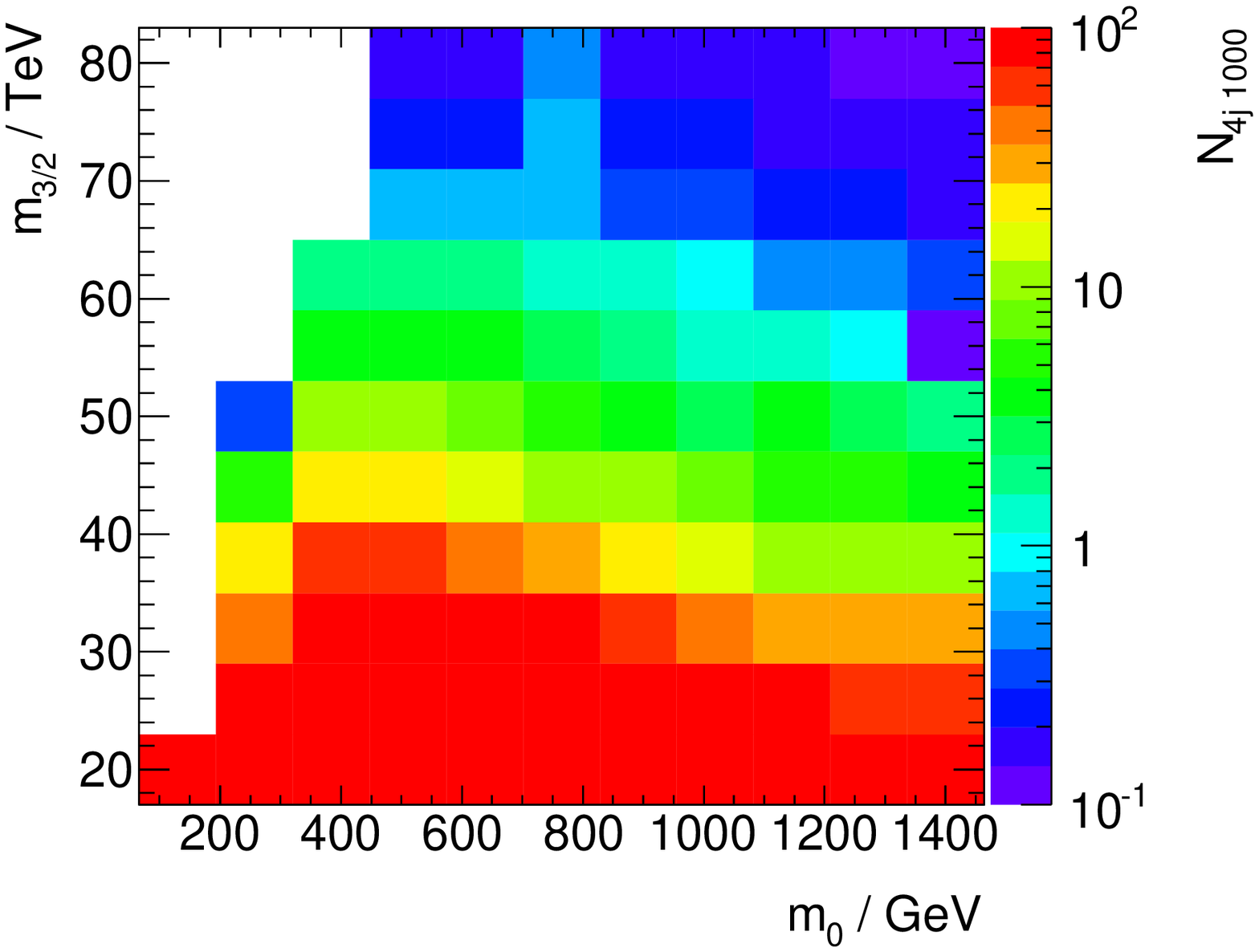}}

\subfigure[high mass]{\includegraphics[width=0.49\textwidth]{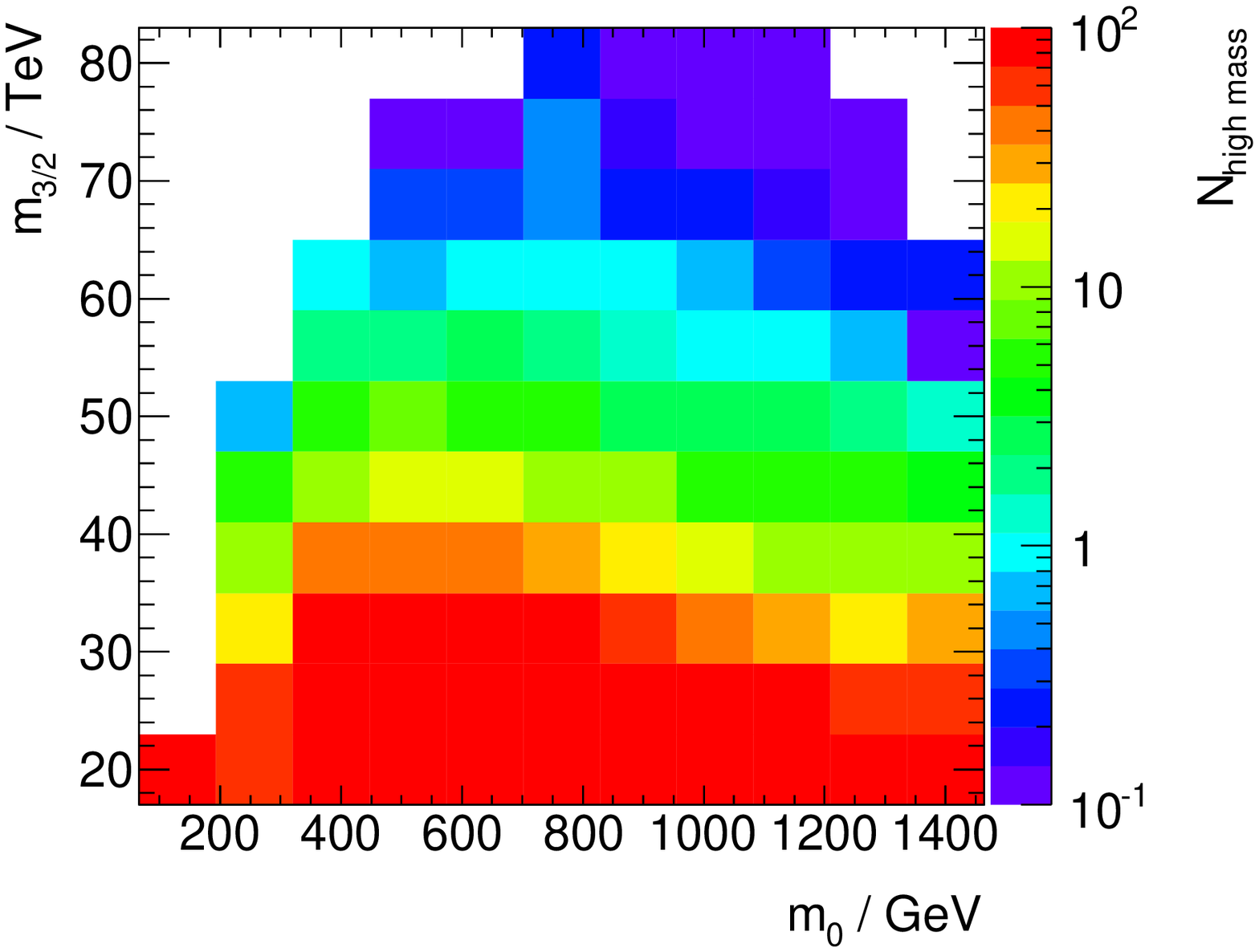}}

\caption{Expected signal yield with 1.04~\ifb{} of ATLAS signal selections in
  the mAMSB $\mzero-\mth$ parameter plane. White areas are either theoretically
  inconsistent, or have fewer than 0.1 expected signal events.}
\label{fig:AMSBefficiency2}
\end{center}
\end{figure}
We show the fractions of events with 0, 1 and 2 leptons
across the mAMSB parameter space in Fig.~\ref{fig:AMSBleptons}.
The isolated lepton veto in the zero lepton search does not cut much of the
SUSY signal over much of the parameter space. On the other hand, searches
based on a 1-lepton channel could also be worthy of study, since over roughly
half of the parameter space, over $15\%$ of SUSY events have a hard lepton. 
We show the efficiency of each signal region in the ATLAS 0-lepton search for
mAMSB in Fig~\ref{fig:AMSBefficiency}. 
The ATLAS selections are seen to be reasonably efficient, particularly at 
greater values of \mth{}, with the exception of a diagonal strip in which the
propensity for producing leptons is greater, as shown in
Fig.~\ref{fig:AMSBleptons}b. 
The ATLAS signal yields at 1.04~\ifb{} are plotted in
Fig.~\ref{fig:AMSBefficiency2} for each signal region. These values are used
to compute the exclusion 
limits on mAMSB\@.
We see that the parameter space we have chosen has roughly the right range of
signal yields expected: na\"ively, in the absence of a signal, we would expect
the regions with tens of 
events at the bottom of each plot to be excluded, whereas regions with only
one or less expected signal events should evade exclusion. 

\subsection{Exclusion Limits in mAMSB}
\begin{figure}[ht]
\begin{center}
\includegraphics[width=0.7 \textwidth]{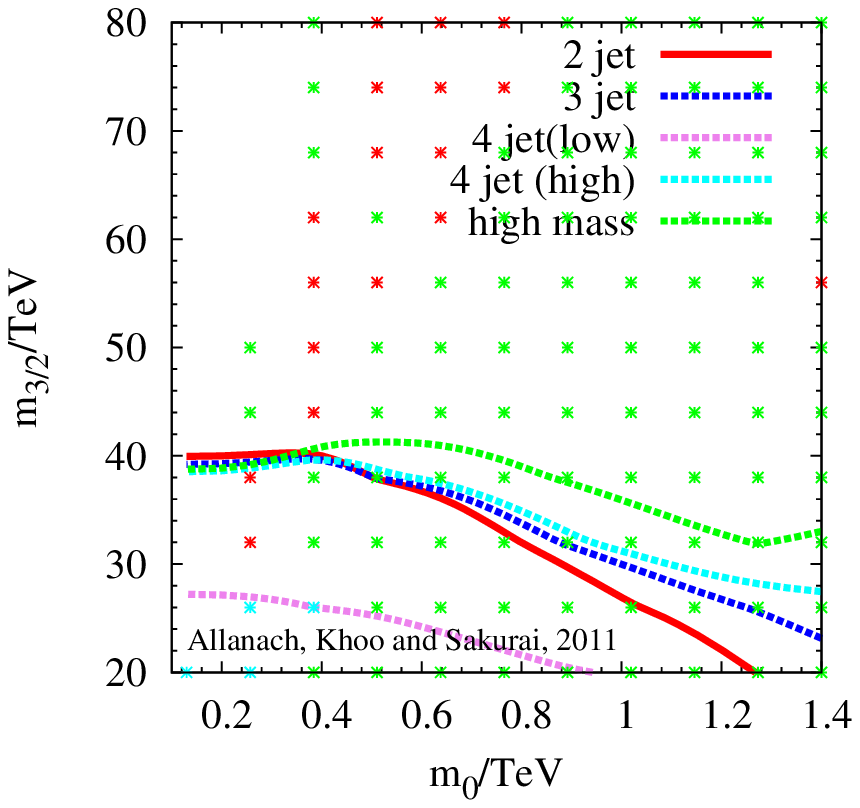}
\end{center}
\caption{ATLAS mAMSB exclusion from the 1.04 fb$^{-1}$ 0-lepton
  search, for   $\tan \beta=10$ 
  and $\mu>0$ for each signal region. 
The region under each line is excluded at the 95$\%$ confidence level for each
individual signal region, labelled by the key and detailed in
Table~\protect\ref{tab:atlascuts}. 
The asterisks in the background display
which signal region is expected to be the most sensitive  at various points in parameter space. The white region in
the upper left hand side of the plot is theoretically disfavoured due to the
presence of negative mass squared sleptons (`tachyons'). 
\label{fig:amsbreach}
}
\end{figure}

We show the 95$\%$ confidence level excluded regions for each signal region 
on the $m_{3/2}-m_0$ parameter space in Fig.~\ref{fig:amsbreach}. We have used
the same systematic errors for each signal region as found in the CMSSM
search in section~\ref{sec:val}. This is another approximation: any variation
of signal systematics  
between the CMSSM and mAMSB is neglected. We expect this approximation to be
good because we obtained a reasonable CMSSM 95$\%$ exclusion limit across the
parameter space, where the sparticle masses are widely varying. 
The most sensitive search regions in mAMSB are the 2-jet
region at low $m_0$ and high $m_{3/2}$ and
the high mass region at large $m_0$. These are the same two regions that are
found to be the most sensitive in the CMSSM\@. 
In the figure, we display a coloured asterisk which
labels which signal region is expected to be the most sensitive at each of our
parameter space grid points. We see that near the exclusion contour, 
for low $m_0$
the 2-jet 
region is expected to be most sensitive,
then for intermediate $m_0 \sim 500$ GeV, the 4-jet high mass region is, and
for $m_0>600$ 
GeV, the high mass region is expected to be the most sensitive.

\subsection{Combination of Signal Regions}

\begin{figure}[ht]
\begin{center}
\includegraphics[width=0.7 \textwidth]{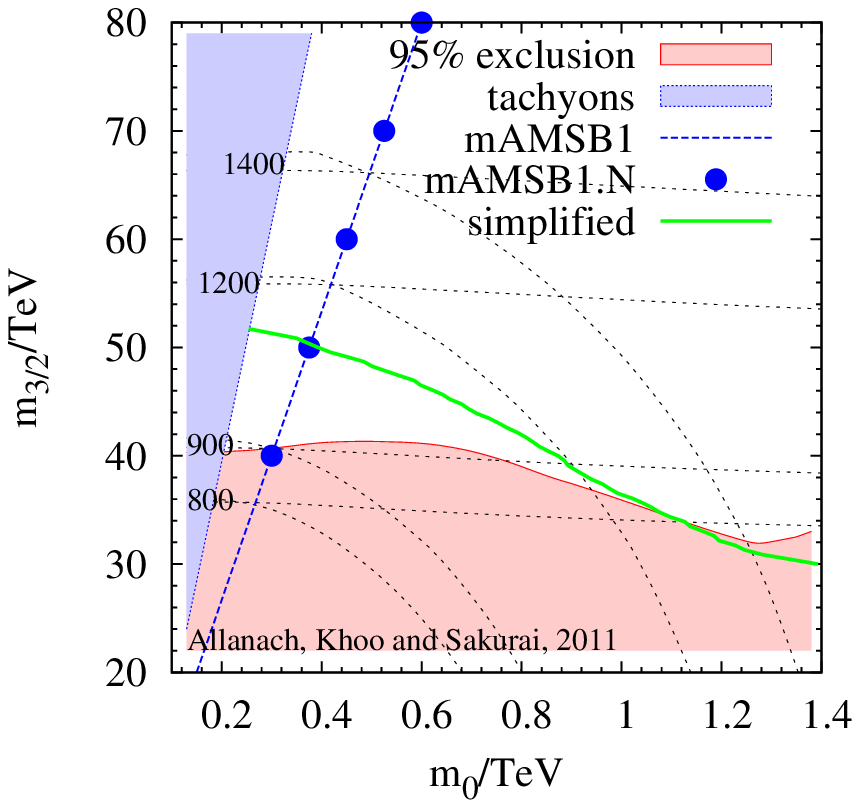}
\end{center}
\caption{ATLAS mAMSB exclusion from the 1.04 fb$^{-1}$ 0-lepton search, for $\tan \beta=10$
  and $\mu>0$, with signal regions combined.
The coloured region is excluded at the 95$\%$ confidence level. The
black dashed lines show equal contours 
of gluino mass (almost horizontal lines) and squark mass (arcs) 
according to the label on the left-hand side of the figure, in units of GeV.
We also show the benchmark mAMSB line and points defined in
Ref.~\cite{AbdusSalam:2011fc} and the simplified model approximation.
\label{fig:amsbtot}
}
\end{figure}
We combine the excluded regions shown in Fig.~\ref{fig:amsbreach} to produce
a single limit contour on the $\mzero-\mth$ plane. In principle, such a
combination 
ought to be carried out independently of the data used to set limits, e.g.\ by
following 
ATLAS' procedure of excluding based on the best \emph{expected} sensitivity of 
each signal region. As can be seen in Fig.~\ref{fig:amsbreach}, this
statistically correct procedure
is approximated in the vicinity of the 95$\%$ exclusion contour by taking
the union of the excluded regions. 
This approximation
sacrifices perfect frequentist coverage  at the point where 
the different most sensitive signal regions cross, in between our grid points
(admittedly, perfect coverage is already abandoned by the use
of the $CL_s$ convention \cite{Read:2000ru}).
As ATLAS observes no significant fluctuations away from the background-only
expectation, our simplified statistical combination (SSC) should deviate only
slightly 
from the ideal coverage. Thus, the SSC should be unobjectionable
to all but the most discerning of frequentist statisticians and US
Congresspersons.  
Experimental collaborations could prevent such abuses as these by including
additional 
information (i.e.\ the expected model-independent limits) in future results.
The combined 95$\%$ CL
exclusion limit thus calculated is shown in Fig.~\ref{fig:amsbtot}, and is the
focal result of our work. 
We also plot on the figure, the approximate exclusion one would obtain
if one naively took the ATLAS 0 lepton simplified model results, in which the
only sparticles are gluinos, squarks and a massless
neutralino~\cite{Alves:2011wf}. The region  
underneath the curve is excluded to 95$\%$ confidence, in the $CL_s$ scheme. 
The trajectory of this curve has squark and gluino masses on the 95$\%$
confidence level contour in the simplified model interpretation of the
0-lepton results~\cite{newATLAS}. We see that over most of the plane, the 
simplified model {\em over estimates} the search reach. This is because the
simplified model assumes 100$\%$ branching ratios into jets, whereas mAMSB has
decays into leptons, reducing the efficiency. Overall, we see that a naive
interpretation of the simplified models would give a poor approximation to the
exclusion\footnote{Nevertheless, it gives a rough ball-park estimate, and
  covers other models than our particular constrained one.}.

When phrased in terms of squark and gluino masses,
mAMSB is {\em less} constrained than the CMSSM. This is due to less
efficient kinematic cuts, as described in Section~\ref{sec:comp}. 
The two most important effects are the lepton veto (18$\%$ less efficient in
our test mAMSB model than the CMSSM model) and a softer \meff{} distribution
in mAMSB (30$\%$ less efficient in the mAMSB test model for the most sensitive
signal region). 
We also display the mAMSB1 benchmark line and points mAMSB1.N (N from 1 to 5,
increasing upwards) from
Ref.~\cite{AbdusSalam:2011fc} in the figure. It is clear that the 1 fb$^{-1}$
data already rule 
out the first point mAMSB1.1 to 95$\%$ C.L., leaving mAMSB1.2 as the next
lightest non-excluded point for study.

\section{Summary and Conclusions \label{sec:summ}}

Recent LHC searches in 1~\ifb{} of integrated luminosity in the jets,
missing transverse momentum and no leptons channel
have seen no
evidence for supersymmetry. These data have been interpreted by the experiments
in terms of the CMSSM and simplified models only. Here, we interpret the data
in terms of an exclusion on the parameter space of mAMSB\@. We also investigate
the mAMSB signals, comparing to the CMSSM in order to see how different or
similar the signal events are expected to be.  
We also wish to examine
how different the exclusion is when phrased in, for example the squark and
gluino masses. 
In fact, mAMSB and the CMSSM have rather similar signal events, although
there is some difference in jet multiplicities and $\ptmiss/\meff$. The
ATLAS 0-lepton cuts involving the latter ratio are slightly more efficient for
our mAMSB test model than for the CMSSM one. 
Because the SUSY events are so similar, there is no reason to  radically
change the cuts in the ATLAS selection. 
The very recent CMS results~\cite{alphaTCMS, mt2CMS, mhtCMS} are slightly more
constraining than those from ATLAS~\cite{newATLAS} in certain regions of the
CMSSM parameter 
space, and so their inclusion to provide combined search limits will be an
interesting exercise for the future. 

We display our summary 95\% C.L. exclusion contour in
Fig.~\ref{fig:amsbtot}, which is the focal result of this paper.  
The final combined 95\% C.L. exclusion limit in mAMSB, at the equal
squark-gluino mass limit, is 900 GeV: a little smaller
than 950 GeV in the 
CMSSM\@. 
Our determination should be accurate to around 30 GeV; 
a more accurate determination would require a dedicated ATLAS analysis, which 
we heartily advocate. 

\acknowledgments
This work has been partially supported by STFC\@. BCA thanks the IPPP for an
associateship. 
TJK is supported by a
Dr. Herchel Smith Fellowship from Williams College. 
KS is supported in part by YLC (Young Leaders Cultivation) program in Nagoya
University. 
We thank A Barr and 
the Cambridge SUSY working group for discussions and P Richardson and D
Grellscheid for their help in massaging {\tt HERWIG++} to accept mAMSB input. 
\bibliographystyle{JHEP}
\bibliography{a}

\end{document}